\documentclass[showpacs,aps,prd,reprint,superscriptaddress,nofootinbib,longbibliography]{revtex4-1}
\usepackage[colorlinks=true, pdfstartview=FitV, bookmarks=true, bookmarksnumbered=true, breaklinks]{hyperref}
\usepackage[dvipdfmx]{graphicx}
\usepackage{amsmath,amssymb,bm,color,longtable,mathrsfs,amsfonts}

\newcommand{\lag}{\mathcal{L}}

\definecolor{DeepPink2}{rgb}{0.932,0.07,0.536}
\definecolor{RoyalBlue1}{rgb}{0.284,0.464,1}
\definecolor{SpringGreen3}{rgb}{0,0.804,0.4}
\definecolor{DarkPastelGreen}{rgb}{0.01,0.69,0.28}
\hypersetup{linkcolor=RoyalBlue1, citecolor=DeepPink2, urlcolor=DarkPastelGreen}

\begin{document}
\begin{flushright}
\end{flushright}

\title{
Strong decays of singly heavy baryons from a chiral effective theory of diquarks
}

\author{Yonghee~Kim}
\email[]{kimu.ryonhi@phys.kyushu-u.ac.jp}
\affiliation{Department of Physics, Kyushu University, Fukuoka 819-0395, Japan}

\author{Makoto~Oka}
\email[]{oka@post.j-parc.jp}
\affiliation{Advanced Science Research Center, Japan Atomic Energy Agency (JAEA), Tokai 319-1195, Japan}
\affiliation{Nishina Center for Accelerator-Based Science, RIKEN, Wako 351-0198, Japan}

\author{Daiki~Suenaga}
\email[]{daiki.suenaga@riken.jp}
\affiliation{Strangeness Nuclear Physics Laboratory, RIKEN Nishina Center, Wako 351-0198, Japan}
\affiliation{Research Center for Nuclear Physics (RCNP), Osaka University, Osaka 567-0047, Japan}

\author{Kei~Suzuki}
\email[]{{k.suzuki.2010@th.phys.titech.ac.jp}}
\affiliation{Advanced Science Research Center, Japan Atomic Energy Agency (JAEA), Tokai 319-1195, Japan}

\date{\today}

\begin{abstract}
A chiral effective theory of scalar and vector diquarks is formulated, which is based on $SU(3)_R\times SU(3)_L$ chiral symmetry and includes interactions between scalar and vector diquarks with one or two mesons.
We find that the diquark interaction term with two mesons breaks the $U(1)_A$ and flavor $SU(3)$ symmetries.
To determine the coupling constants of the interaction Lagrangians, we investigate one-pion emission decays of singly heavy baryons $Qqq$ ($Q=c, b$ and $q=u,d,s$), where baryons are regarded as diquark--heavy-quark two-body systems.
Using this model, we present predictions of the unobserved decay widths of singly heavy baryons. We also study the change of masses and strong decay widths of singly heavy baryons under partial restoration of chiral symmetry.  
\end{abstract}
\maketitle
\section{Introduction}\label{Sec_1}
The diquark is a color nonsinglet state made of two quarks \cite{GellMann:1964nj, Ida:1966ev, Lichtenberg:1967zz, Lichtenberg:1967, Souza:1967rms, Lichtenberg:1968zz, Carroll:1969ty, Lichtenberg:1981pp, RevModPhys.65.1199, JAFFE20051} and has been realized to play important roles in hadron spectroscopy as a substructure of baryons and exotic hadrons. Diquarks also appear in the color superconducting phase in high-density QCD~\cite{Alford:1997zt, Rapp:1997zu}.
The properties of diquark itself, such as the mass and the size, have been studied by lattice QCD simulations~\cite{Hess:1998sd, Orginos:2005vr, Alexandrou:2006cq, Babich:2007ah, DeGrand:2007vu, Green:2010vc, ChinQCD, Watanabe:2021oyv, Watanabe:2021nwe, Francis:2021vrr, Francis:2022fdj}.

Since the diquark is a colorful object, it cannot be observed directly in experiments. To investigate the features of diquarks with making up for this point, singly heavy baryons ($Qqq$) are useful. They are composed of one heavy (charm or bottom) quark ($Q=c, b$) and two light (up, down, or strange) quarks ($q=u, d, s$), so that the two light quarks ($qq$) may be described as the diquark inside these baryons \cite{Lichtenberg:1975ap, Lichtenberg:1982jp, Lichtenberg:1982nm, Fleck:1988vm, Ebert:1995fp, Ebert:2007nw, Kim:2011ut, Ebert:2011kk, Chen:2014nyo, Jido:2016yuv, Kumakawa:2017ffl, CETdiquark, scalar, Dmitrasinovic:2020wye, Kawakami:2020sxd, Suenaga:2021qri, Vdiquark, Suenaga:2022ajn}. 

 Another important aspect of hadron spectroscopy is the chiral symmetry and its spontaneous symmetry breaking in the low-energy regime of QCD. In our previous works, an effective Lagrangian based on the chiral $SU(3)_R \times SU(3)_L$ symmetry for diquarks was constructed. In Refs.~\cite{CETdiquark, scalar}, the chiral effective model of scalar and pseudoscalar diquarks with spin $0$ is proposed and is employed to investigate the mass spectra of singly heavy baryons with the diquark--heavy-quark potential model. Here we derived the mass formulas for the diquarks and discovered the inverse mass hierarchy of diquark masses: The nonstrange pseudoscalar diquark becomes heavier than the singly strange pseudoscalar diquark [$M(ud, 0^-) > M(ds/su, 0^-)$] due to the $U(1)_A$ anomaly~\cite{tHooft:1976snw, tHooft:1986ooh}. Besides, in Ref.~\cite{Vdiquark}, we constructed the chiral effective model of vector and axial-vector diquarks with spin $1$ and updated our numerical results of the spectrum of singly heavy baryons with a renewal of potential models. We obtained the mass formula for the axial-vector diquarks [$M^2(uu/ud/dd, 1^+) + M^2(ss, 1^+) = 2M^2(ds/su, 1^+)$], which is a generalization of the Gell-Mann--Okubo mass formula \cite{gomass1, gomass2}.\footnote{From the viewpoints of diquarks in a chiral framework, there are some other works. For light mesons, see Refs. \cite{Fariborz:2005gm, Giacosa:2006tf, Fariborz:2007ai}. For an application to light baryons, see Ref. \cite{Olbrich:2015gln}.}

One of the characteristic points of these works is applying mass formulas of diquarks to the spectrum of singly heavy baryons. Here scalar and vector diquarks are considered independently and interactions between these diquarks are neglected because they are expected to be irrelevant to the diquark mass formulas. Instead of this, however, such an interaction plays the main role in strong decays of singly heavy baryons~\cite{Isgur:1991wq, Yan:1992gz, Cho:1992gg, Cho:1994vg, Rosner:1995yu, Albertus:2005zy, Cheng:2006dk, Zhong:2007gp, Yasui:2014cwa, Nagahiro, Chen:2016iyi, Can:2016ksz, Chen:2017sci, Wang:2017hej, Arifi:2017sac, Wang:2017kfr, Yao:2018jmc, Kawakami:2018olq, Lu:2018utx, Arifi:2018yhr, Wang:2018fjm, Arifi:2019lny, Kawakami:2019hpp, Cui:2019dzj, Nieves:2019nol, Wang:2019uaj, Yang:2019cvw, Lu:2019rtg, Chen:2020mpy, Azizi:2020tgh, Yang:2020zrh, Arifi:2020ezz, Yang:2020zjl, Azizi:2020azq, Arifi:2021orx, Arifi:2021wdf, Yang:2021lce, Gong:2021jkb, Kakadiya:2021jtv, Suenaga:2022ajn, Suh:2022ean, Garcia:2022zrf, Yanga:2022oog, Yu:2022ymb}.

In this paper, we extend our approach~\cite{CETdiquark, scalar,Vdiquark} to interactions between the scalar and vector diquarks, which satisfies the chiral $SU(3)_R\times SU(3)_L$ symmetry.
We determine the coupling constants for these interactions from the experimentally known strong decay widths of singly heavy baryons and also investigate the effect of chiral symmetry restoration. 

This paper is organized as follows.
In Sec.~\ref{Sec_2}, we formulate a chiral effective model including interactions between scalar and vector diquarks.
In Sec.~\ref{Sec_3}, we investigate one-pion emission decays of singly heavy baryons based on a diquark--heavy-quark description and determine the coupling constants of the interaction Lagrangian. Besides, modifications of decay widths of the baryons triggered by the partial restoration of chiral symmetry are demonstrated.
Finally, Sec.~\ref{Sec_4} is for the conclusion of the present work.

\section{Chiral effecive model of diquarks} \label{Sec_2}
\label{sec:Diquark}

Main purpose of this work is to study strong decays of singly heavy baryons from the diquark--heavy-quark description based on chiral symmetry. In this framework, interactions between singly heavy baryons and light mesons are determined by chiral dynamics of the diquarks and light mesons, where the remaining heavy quark simply serves as a spectator. To this end, in this section we introduce a chiral effective Lagrangian describing interactions between diquarks and light mesons. For the diquarks, in particular, we include the vector (V) and axial-vector (A) diquarks as well as the scalar (S) and pseudo-scalar (P) ones.

\subsection{Diquark operators}
Within a simple description, singly heavy baryons are composed of one heavy quark and one diquark belonging to the color $3$ and $\bar{3}$ representations, respectively. Here, we briefly explain the properties of color $\bar{3}$ diquarks and their structures based on chiral symmetry.

 \begin{table}[b]
  \centering
    \caption{Local diquark operators belonging to color $\bar{3}$.}      
  \begin{tabular}{ l | c | c | c | c  } \hline\hline
    Diquark & Operator & $J^P$ & Color & Flavor   \\ \hline \hline
 S (scalar) & $(q^T C\gamma^5 q)^{\overline{3}}_{A}$ &$0^+$&$\overline{3}$&$\overline{3}$\\ 
 P (pseudoscalar)& $(q^T Cq)^{\overline{3}}_{A}$ &
                                                                                    $0^-$&$\overline{3}$&$\overline{3}$\\ 
 V (vector) & $(q^T C\gamma^{\mu} \gamma^5 q)^{\overline{3}}_{A}$ &
                                                                                         $1^-$&$\overline{3}$&$\overline{3}$\\    
 A (axial-vector) & $(q^T C\gamma^{\mu} q)^{\overline{3}}_{S}$ &
                                                                                         $1^+$&$\overline{3}$& 6 \\ \hline \hline
  \end{tabular}  
  \label{4diquarks}
  \end{table}

In Table~\ref{4diquarks}, we summarize interpolating fields (or operators) of the diquarks studied in this work and their quantum numbers.\footnote{Note that the S and A diquarks are often referred to as the ``good'' and ``bad'' diquarks, respectively~\cite{Jaffe:2005zz, Kopeliovich:2005hs, Kopeliovich:2008dj, Chen:2009tm}.} In this table, $C=i \gamma^0 \gamma^2$ is the charge conjugation Dirac matrix. The subscript ``$A$'' and ``$S$'' stand for the antisymmetric and symmetric combinations in flavor indices, respectively, which are shown in the rightmost column.
The superscript $\bar{3}$ means that the diquarks are antisymmetric in color indices.
Their spin and parities are represented by $J^P$.

 \begin{table}[tb]
  \centering
    \caption{Local diquark operators in the chiral basis for diquarks in Table \ref{4diquarks}.}   
  \begin{tabular}{ l | c | c | c  } \hline\hline
   Chiral operator & Spin & Color & Chiral \\ \hline \hline
 $d^a_{R,i} = \epsilon^{abc}\epsilon_{ijk}(q^{bT}_{R,j} Cq^c_{R,k})$ &
 										$0$&$\overline{3}$&$(\overline{3}, 1)$\\ 
 $d^a_{L,i} = \epsilon^{abc}\epsilon_{ijk}(q^{bT}_{L,j} Cq^c_{L,k})$ &
                                                                                    $0$&$\overline{3}$&$(1, \overline{3})$\\ 
 $d^{a,\mu}_{ij} = \epsilon^{abc}(q^{bT}_{L,i} C \gamma^\mu q^c_{R,j})$ &
                                                                                         $1$&$\overline{3}$&$(3,3)$\\  \hline \hline
  \end{tabular}  
  \label{chiraldiquark}
  \end{table}
  
The diquarks listed in Table~\ref{4diquarks} are parity eigenstates which are useful to see connections with physical states of the singly heavy baryons. To see diquarks from the aspects of chiral symmetry, it is useful to classify them in terms of the left-handed and right-handed quarks defined by $q_{R,i}^a= P_Rq_i^a$ and $q_{L,i}=P_Lq_i^a$ with the chiral projection operators $P_{R,L}\equiv (1 \pm \gamma^5)/2$. Here, $a$ and $i$ denote the color and flavor indices, respectively. In such chiral basis, the four diquark operators in Table~\ref{4diquarks} are decomposed to three chiral diquark operators given in Table~\ref{chiraldiquark}. From the definition, 
$q^a_{R,i}$ and $q^a_{L,i}$ belong to the $(3, 1)$ and $(1, 3)$ representations of $SU(3)_R\times SU(3)_L$ chiral symmetry, respectively,  and accordingly the chiral representations of the diquarks are determined as in the table.

Below, we show the chiral and parity transformations of the diquarks in Table~\ref{chiraldiquark}. First, since $q^a_{R,i}$ and $q^a_{L,i}$ belong to $(3, 1)$ and $(1, 3)$, respectively, they transform under $SU(3)_R\times SU(3)_L$ chiral symmetry as $q^a_{R,i} \rightarrow (U_R)_{ij} q^a_{R,j}$ and $q^a_{L,i} \rightarrow (U_L)_{ij} q^a_{L,j}$, with $U_R\in SU(3)_R$ and $U_L\in SU(3)_L$. Thus, one can easily see that the chiral diquarks are transformed as~\cite{CETdiquark}
\begin{equation}
\begin{split}
&{\rm Ch} \ : \ d^a_{R, i} \rightarrow d^a_{R_j} (U^\dag_R)_{ji},~d^a_{L, i} \rightarrow d^a_{L_j} (U^\dag_L)_{ji},\\
&{\rm Ch} \ : \  d^{a,\mu}_{ij} \rightarrow U_{L,i k} ~d^{a,\mu}_{km} ~U^T_{R,m j}.
\end{split}
\label{cvdt}
\end{equation}
Next, it is well known that the spatial inversion of the left-handed or right-haded quark is given by $q^a_{R/L, i} (t, \vec{x}) \rightarrow \gamma_0 q^a_{L/R, i} (t, -\vec{x})$. From this formula, the parity transformation of the chiral diquarks reads
\begin{equation}
\begin{split}
&\mathcal{P}\ :\  d^a_{R/L,i} \rightarrow -d^a_{L/R,i},\\
&\mathcal{P}\ :\ d^{a,\mu}_{ij} \rightarrow -d^a_{\mu,ji}.
\end{split}
\label{parity}
\end{equation}
These transformation properties are important for constructing a chiral Lagrangian of the diquarks. Besides, the parity transformation law in Eq.~(\ref{parity}) enables us to express the parity-eigenstate diquark operators in Table \ref{4diquarks} in terms of the chiral operators in Table~\ref{chiraldiquark} as
\begin{eqnarray}
&&S^a_i=\frac{1}{\sqrt{2}}(d^a_{R,i}-d^a_{L,i})
	     =\epsilon_{abc}\epsilon_{ijk}\frac{1}{\sqrt{2}}(q^{bT}_{j} C \gamma^5 q^c_k),~~
\label{Sdiquark}\\
&&P^a_i=\frac{1}{\sqrt{2}}(d^a_{R,i}+d^a_{L,i})
	     =\epsilon_{abc}\epsilon_{ijk}\frac{1}{\sqrt{2}}(q^{bT}_{j} C q^c_k),~~	
\label{Pdiquark}\\
&&V^{a,\mu}_{ij}=\frac{1}{\sqrt{2}}(d^{a,\mu}_{ij}-d^{a,\mu}_{ji})
		     =\epsilon_{abc}\frac{1}{\sqrt{2}}(q_i^{bT}C\gamma^{\mu}\gamma^{5}q_j^c),~~
 \label{Vdiquark}\\
&&A^{a,\mu}_{ij}=\frac{1}{\sqrt{2}}(d^{a,\mu}_{ij}+d^{a,\mu}_{ji})
		     =\epsilon_{abc}\frac{1}{\sqrt{2}}(q_i^{bT}C\gamma^{\mu}q_j^c).~~
 \label{Adiquark}
\end{eqnarray}
From these expressions, we find that S and P diquarks are chiral partners belonging to the chiral ($\bar{3},1$)$+$($1,\bar{3}$) representation, while V and A diquarks are chiral partners belonging to the chiral ($3,3$) representation.

\begin{figure*}[t]
\centering
  \includegraphics[clip,width=2.0\columnwidth]{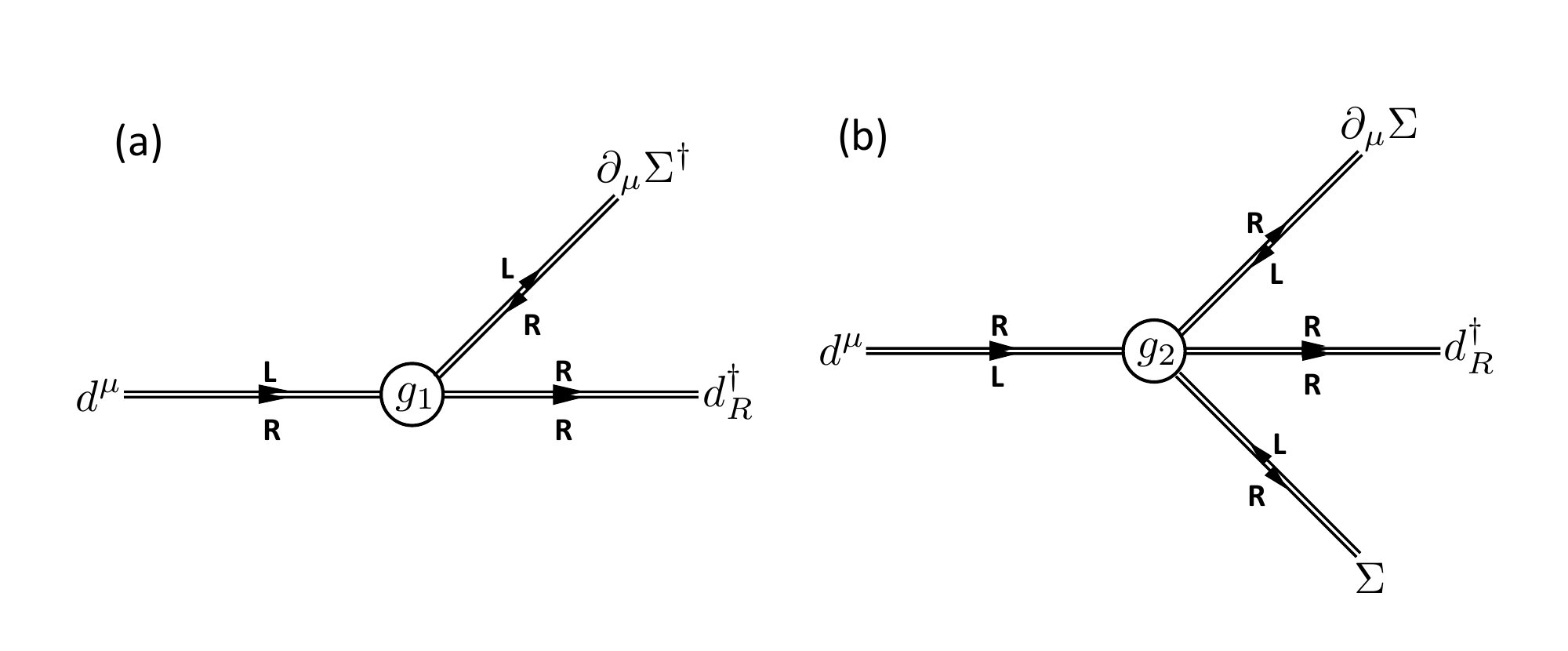}
  \caption{Schematic pictures for the effective interaction Lagrangian $\lag_{SV}$ [Eq.~(\ref{SVall})] between diquarks and mesons.
The left figure (a) and right figure (b) describe $\lag_{SV}^{(1)}$ [Eq.~(\ref{SV1})] and $\lag_{SV}^{(2)}$ [Eq.~(\ref{SV2})], respectively.
Arrows with ``R" or ``L" represent the right-handed or left-handed quark lines.
} 
  \label{quarkline}
\end{figure*}

\subsection{Chiral effective Lagrangian}
\label{sec:ChiralLag}

In this subsection, we present interaction Lagrangian including the diquarks and light mesons based on $SU(3)_R\times SU(3)_L$ chiral symmetry.

According to Ref.~\cite{Vdiquark}, the chiral effective Lagrangian for diquarks in Table~\ref{chiraldiquark} and light mesons are expressed within the linear sigma model as
\begin{eqnarray}
\lag=\frac{1}{4}{\rm Tr}[\partial^\mu \Sigma^\dag \partial_\mu \Sigma]-V(\Sigma)
+\mathcal{L}_S + \mathcal{L}_V +\mathcal{L}_{SV}.
\label{lagall}
\end{eqnarray}
In the first and second terms, $\Sigma_{ij}$ is the meson nonet composed of the scalar $\sigma_{ij}$ and pseudoscalar $\pi_{ij}$ mesons \cite{GellMann:1960np, levy1967currents}. The latter mesons are regarded as the Nambu-Goldstone (NG) bosons in association with the breakdown of $SU(3)_R\times SU(3)_L$ chiral symmetry triggered by the instability of the potential $V(\Sigma)$ at $\Sigma=0$, where $V(\Sigma)$ is the potential terms of the meson nonet.
Since the meson nonet $\Sigma_{ij}=\bar{q}_{R,j}q_{L,i}$ belongs to chiral ($\bar{3},3$) representation, its chiral transformation is given as
\begin{eqnarray}
\Sigma_{ij}=\sigma_{ij}+i\pi_{ij}\rightarrow U_{L,ik}\Sigma_{km}U_{R,mj}^\dag .
\label{LSM}
\end{eqnarray}
The third and forth terms in Eq.~(\ref{lagall}) represent the kinetic and mass terms of the diquarks, where $\lag_S$ includes spin 0 (S and P) diquarks while $\lag_{V}$ includes spin 1 (V and A) diquarks.
For their explicit expressions, see Refs.~\cite{CETdiquark, Vdiquark}.

The last term of chiral effective Lagrangian in Eq.~(\ref{lagall}), $\lag_{SV}$, represents couplings between the diquarks with different spins mediated by the meson nonet. In this work, we propose the following two terms:
\begin{eqnarray}
&&\lag_{SV}=\lag_{SV}^{(1)}+\lag_{SV}^{(2)}, 
\label{SVall}
\end{eqnarray}
where
\begin{eqnarray}
&&\lag_{SV}^{(1)}
=g_1\epsilon_{ijk}
\left[
d^\mu_{ni} (\partial_\mu \Sigma^\dag)_{jn} d^\dag_{R,k}
+d^\mu_{in} (\partial_\mu \Sigma)_{jn} d^\dag_{L,k}
 \right],~~~~
 \label{SV1} 
 \end{eqnarray}
 and
 \begin{eqnarray}
&&\lag^{(2)}_{SV}
=\frac{g_2}{f_\pi}\epsilon_{ijk}
\left[
d^\mu_{in} \{\Sigma_{jn}(\partial_\mu \Sigma)_{km}-(\partial_\mu \Sigma)_{jn}\Sigma_{km}\} d^\dag_{R,m}\right. \nonumber \\
&&\hspace{0.9cm}\left.
+d^\mu_{ni} \{\Sigma^\dag_{jn}(\partial_\mu \Sigma^\dag)_{km}-(\partial_\mu \Sigma^\dag)_{jn}\Sigma^\dag_{km}\} d^\dag_{L,m} 
 \right].~~~~
 \label{SV2}
\end{eqnarray}
The coefficients $g_1$ and $g_2$ are dimensionless coupling constants, and $f_\pi$ in Eq. (\ref{SV2}) is the pion decay constant. All the color indices are implicitly contracted. In constructing the Lagrangian~(\ref{SVall}) we have taken into account the contributions up to ${\cal O}(\Sigma^2)$ so as to satisfy chiral symmetry in Eqs.~(\ref{cvdt}) and~(\ref{LSM}). 

Since the spatial inversion of $\Sigma$ is expressed as $\mathcal{P}: \Sigma(t,\vec{x}) \rightarrow \Sigma^\dag(t,-\vec{x})$, the Lagrangians in Eqs.~(\ref{SV1}) and (\ref{SV2}) are invariant under parity transformation as well as chiral transformation. However, under the axial $U(1)_A$ transformation, only $\lag_{SV}^{(1)}$ keeps the symmetry while $\lag_{SV}^{(2)}$ breaks it. That is, the latter is responsible for the $U(1)_A$ anomaly effects, 
which may be caused by instantons (a topologically nontrivial configuration of gluon fields) \cite{instanton}.

In order to diagrammatically understand the $U(1)_A$ anomaly effects, in Fig. \ref{quarkline}, we draw schematic pictures of the vertices and the quark lines: In the left figure (a) for $\lag_{SV}^{(1)}$, all the right-handed or left-handed quark lines, denoted by ``R'' or ``L'', is conserved through the interaction vertex.
On the other hand, in the right figure (b) for $\lag_{SV}^{(2)}$, the quark lines are not conserved.

\subsection{Explicit chiral symmetry breaking}

In the vacuum where $SU(3)_R\times SU(3)_L$ chiral symmetry is spontaneously broken,  the vacuum expectation values (VEVs) of the meson nonet $\Sigma$ is nonzero: $\langle\Sigma\rangle \neq 0$. As a result, the $\Sigma$ with no derivatives in Eq.~(\ref{SV2}) can be replaced by its VEV $\langle\Sigma\rangle$, and couplings describing
one meson emission decays of the diquarks are obtained from Eq.~(\ref{SV2}) as well as Eq.~(\ref{SV1}). In such treatment, effects from the violation of $SU(3)_R\times SU(3)_L$ chiral symmetry due to a mass of strange quark cannot be ignored.
In this subsection, we explain our method to incorporate such explicit chiral symmetry breaking (ECSB) effects into the Lagrangian~(\ref{SVall}).

When taking into account the current quark masses, 
within the linear sigma model, the mass matrix $\mathcal{M}_{\rm eff}$ of constituent quarks can be expressed as
\begin{eqnarray}
&&\mathcal{M}_{\rm eff}=\mathcal{M}+g_s\langle\Sigma\rangle ,
\label{MM}
\end{eqnarray}
where $\mathcal{M}={\rm diag}(m_u, m_d, m_s)$ denotes the current quark mass matrix, and $\langle\Sigma\rangle$ is the VEV of $\Sigma$ with $g_s$ the quark-meson coupling constant. As for the current quark masses, the lattice QCD simulations suggest that $m_s= 93.1\, {\rm MeV}$ while $m_u\approx m_d\approx 2-5\, {\rm MeV}$, and hence we take $m_u=m_d=0$ as a good approximation and assume $SU(2)_I$ isospin symmetry throughout the symmetry breaking. In this case, the VEV of $\Sigma$ must be diagonal as $\langle\Sigma\rangle = {\rm diag}(\langle\sigma_{11}\rangle,\langle\sigma_{22}\rangle,\langle\sigma_{33}\rangle)$ with $\langle\sigma_{11}\rangle = \langle\sigma_{22}\rangle$. These values are determined by the pion and kaon decay constants $f_\pi$ and $f_K$. That is, by evaluating the axial currents from Eq.~(\ref{lagall}), one can find $\langle\sigma_{11}\rangle=\langle\sigma_{22}\rangle=f_\pi = 92.1$ MeV and $\langle\sigma_{33}\rangle=f_s \equiv 2 f_K-f_\pi = 128.1$ MeV, where $f_\pi = 92.1$ MeV and $f_K=110.1$ MeV are from the Particle Data Group (PDG)~\cite{PDG}. As a result, the effective quark mass matrix can be expressed in a simple form as
\begin{equation}
\begin{split}
&\mathcal{M}_{\rm eff} \simeq g_s f_\pi{\rm diag}(1,1,\alpha), \\
&\alpha = \frac{f_s}{f_\pi} \left( 1+\frac{m_s}{g_s f_s} \right) .
\label{effectivequarkmass}
\end{split}
\end{equation}

One of the most useful ways to incorporate the ECSB effects is to replace $\Sigma$ in the Lagrangian~(\ref{lagall}) by the following shifted nonet $\tilde{\Sigma}$~\cite{CETdiquark, Vdiquark}:
 \begin{eqnarray}
 \Sigma \rightarrow\tilde{\Sigma} \equiv \Sigma + \mathcal{M}/g_s,
 \label{Sigmatilde}
\end{eqnarray}
where its VEV is given by 
\begin{eqnarray}
\langle \tilde{\Sigma} \rangle = \mathcal{M}_{\rm eff}/g_s=f_\pi {\rm diag}(1,1,\alpha).
\label{Sigmatildevev}
\end{eqnarray}
By substituting Eqs. (\ref{effectivequarkmass})--(\ref{Sigmatildevev}) into the effective Lagrangian~(\ref{SVall}), we obtain the interaction Lagrangian with the ECSB effects:
\begin{eqnarray}
{\cal L}_{SV} = \lag_{AS\pi} + \lag_{VP\pi}\ , \label{LSVExplicit}
\end{eqnarray}
with
\begin{eqnarray}
i\lag_{AS\pi}
&=(g_1+\alpha g_2)\left[
A^\mu_{1n}(\partial_\mu\pi)_{2n}S^\dag_3-A^\mu_{2n}(\partial_\mu\pi)_{1n}S^\dag_3
\right] \nonumber\\
&\hspace{0.3cm}
+(g_1+g_2)\left[
A^\mu_{2n}(\partial_\mu\pi)_{3n}S^\dag_1-A^\mu_{3n}(\partial_\mu\pi)_{2n}S^\dag_1\right. \nonumber\\
&\hspace{0.6cm}
\left. +A^\mu_{3n}(\partial_\mu\pi)_{1n}S^\dag_2-A^\mu_{1n}(\partial_\mu\pi)_{3n}S^\dag_2
\right], 
\label{lagASpi}
\end{eqnarray}
and
\begin{eqnarray}
i\lag_{VP\pi}
&=-(g_1-\alpha g_2)\left[
V^\mu_{1n}(\partial_\mu\pi)_{2n}P^\dag_3-V^\mu_{2n}(\partial_\mu\pi)_{1n}P^\dag_3
\right]\nonumber\\
&\hspace{0.2cm}
-(g_1-g_2)\left[
V^\mu_{2n}(\partial_\mu\pi)_{3n}P^\dag_1-V^\mu_{3n}(\partial_\mu\pi)_{2n}P^\dag_1 \right. \nonumber \\
&\hspace{1.8cm}
\left.+V^\mu_{3n}(\partial_\mu\pi)_{1n}P^\dag_2-V^\mu_{1n}(\partial_\mu\pi)_{3n}P^\dag_2\right]\nonumber\\
&\hspace{0.4cm}
-(1+\alpha)g_2 \left[V^\mu_{23}(\partial_\mu\pi)_{1n}P^\dag_n+V^\mu_{31}(\partial_\mu\pi)_{2n}P^\dag_n\right] \nonumber \\
&\hspace{0.0cm}
 -2g_2\left[V^\mu_{12}(\partial_\mu\pi)_{3n}P^\dag_n\right].
\label{lagVPpi}
\end{eqnarray}
In order to obtain Eqs.~(\ref{lagASpi}) and~(\ref{lagVPpi}), we have rearranged the $g_1$ and $g_2$ terms in Eqs.~(\ref{SV1}) and~(\ref{SV2}) such that $\lag_{AS\pi}$ and $\lag_{VP\pi}$ contain couplings of the S and A diquarks and those of the P and V diquarks, respectively.
Note that we have omitted interactions mediated by scalar mesons $\sigma_{ij}$ in Eqs.~(\ref{lagASpi}) and~(\ref{lagVPpi}). 

When we decompose the pseudoscalar nonet $\pi_{ij}$ into the singlet and octet parts as
\begin{eqnarray}
\pi_{ij}
&&=\sqrt{\frac{2}{3}}\eta_1\delta_{ij}+\sqrt{2}
\begin{pmatrix}
\frac{\pi^0}{\sqrt{2}}+\frac{\eta_8}{\sqrt{6}}&\pi^+&K^+\\
\pi^-&-\frac{\pi^0}{\sqrt{2}}+\frac{\eta_8}{\sqrt{6}}&K^0\\
K^-&\bar{K}^0&-\frac{2\eta_8}{\sqrt{6}}
\end{pmatrix}\nonumber\\
&& \hspace{0.0cm} \equiv\sqrt{\frac{2}{3}}\pi^{(1)}\delta_{ij}+\pi^{(8)}_{ij},
\label{meson18}
\end{eqnarray}
the flavor structures of Eq.~(\ref{LSVExplicit}) become more transparent. The Lagrangian written in terms of $\pi_{ij}^{(1)}$ and $\pi_{ij}^{(8)}$ is straightforwardly obtained by substituting Eq.~(\ref{meson18}) into Eq.~(\ref{LSVExplicit}), but the resultant expression is lengthy. Hence, here we only comment on its flavor symmetric properties. The resultant Lagrangian indicates that the flavor-singlet pseudoscalar meson $\pi^{(1)}$ does not couple to A and S diquarks. This is because the A and S diquarks belong to flavor $6$ and $\bar{3}$ representations, respectively, so that original flavor symmetry prohibits such couplings. Similarly, both the V and P diquarks belong to flavor $\bar{3}$, and thus they couple to both flavor singlet $\pi^{(1)}$ and octet $\pi^{(8)}$ mesons.\footnote{Although the arguments based on $SU(3)$ flavor symmetry is valid only for $\alpha=1$ where original $SU(3)_R\times SU(3)_L$ symmetry is exactly satisfied, it seems that they can be applied to the case for $\alpha\neq1$.}

\section{Decays of $\Sigma_Q$ and $\Xi_Q'$ baryons} \label{Sec_3}

In Sec.~\ref{sec:Diquark}, we have formulated the chiral effective model of diquarks describing their one-pion emission decays. In this section, from the chiral model and the diquark--heavy-quark description, we investigate strong decays of singly heavy baryons and quantify the model using the experimental data. 
In Table~\ref{PDGdata}, we summarize the experimental data of singly heavy baryons from the PDG~\cite{PDG}, which are expected to be in the ground states. 
\begin{table}[t]
  \centering
    \caption{Experimental values of masses and decay widths of singly heavy baryons in units of MeV, taken from the PDG compilation \cite{PDG}. We also use the pion masses as $m_{\pi^\pm}=139.57$ MeV and $m_{\pi^0}=134.97$ MeV.}
  \begin{tabular}{ c  c  c  c  c  } \hline\hline
       \multicolumn{1}{c}{Baryon}
       &\multicolumn{1}{c}{$J^P$}
       &\multicolumn{1}{c}{Mass}
       &\multicolumn{1}{c}{Full width}
       &\multicolumn{1}{c}{Decay mode}    \\
       \multicolumn{1}{c}{}
       &\multicolumn{1}{c}{}
       &\multicolumn{1}{c}{(MeV)}
       &\multicolumn{1}{c}{(MeV)}
       &\multicolumn{1}{c}{}    \\ \hline       
       
       \hline
\multicolumn{1}{c}{$\Lambda_c$}
&\multicolumn{1}{c}{$1/2^+$}
&\multicolumn{1}{c}{$2286.46$}	
&\multicolumn{2}{c}{(No strong decay)} \\ \hline
$\Sigma_c^{++}(2455)$	&$1/2^+$&$2453.97$	
&$1.89$ 	&$\Sigma_c^{++} \rightarrow \Lambda_c \pi^+$\\ 
$\Sigma_c^{+}(2455)$	&$1/2^+$&$2452.9$	
& ($<4.6$) 			&$\Sigma_c^{+} \rightarrow \Lambda_c \pi^0$ \\ 
$\Sigma_c^{0}(2455)$	&$1/2^+$&$2453.75$	
& $1.83$ 	&$\Sigma_c^{0} \rightarrow \Lambda_c \pi^-$\\ \hline
$\Sigma_c^{++}(2520)$	&$3/2^+$&$2518.41$	
& $14.78$&$\Sigma_c^{*++} \rightarrow \Lambda_c \pi^+$\\ 
$\Sigma_c^{+}(2520)$	&$3/2^+$&$2517.5$	
& ($<17$)				&$\Sigma_c^{*+} \rightarrow \Lambda_c \pi^0$\\ 
$\Sigma_c^{0}(2520)$	&$3/2^+$&$2518.48$	
& $15.3$  	&$\Sigma_c^{*0} \rightarrow \Lambda_c \pi^-$\\ \hline
\multicolumn{1}{c}{$\Xi_c^{+}$}			
&\multicolumn{1}{c}{$1/2^+$}
&\multicolumn{1}{c}{$2467.71$}	
&\multicolumn{2}{c}{(No strong decay)} \\ 
\multicolumn{1}{c}{$\Xi_c^{0}$}
&\multicolumn{1}{c}{$1/2^+$}
&\multicolumn{1}{c}{$2470.44$}	
&\multicolumn{2}{c}{(No strong decay)} \\ \hline
\multicolumn{1}{c}{$\Xi_c^{'+}$}
&\multicolumn{1}{c}{$1/2^+$}
&\multicolumn{1}{c}{$2578.2$}	
&\multicolumn{2}{c}{(No strong decay)} \\ 
\multicolumn{1}{c}{$\Xi_c^{'0}$}
&\multicolumn{1}{c}{$1/2^+$}
&\multicolumn{1}{c}{$2578.7$}	
&\multicolumn{2}{c}{(No strong decay)} \\ \hline
$\Xi_c^{+}(2645)$		&$3/2^+$&$2645.10$	
&$2.14$  	&$\Xi_c^{'*+} \rightarrow \Xi_c^{0(+)} \pi^{+(0)}$\\ 
$\Xi_c^{0}(2645)$		&$3/2^+$&$2646.16$	
&$2.35$ 	&$\Xi_c^{'*0} \rightarrow \Xi_c^{+(0)} \pi^{-(0)}$\\ \hline

\hline
\multicolumn{1}{c}{$\Lambda_b$}
&\multicolumn{1}{c}{$1/2^+$}
&\multicolumn{1}{c}{$5619.60$}	
&\multicolumn{2}{c}{(No strong decay)} \\ \hline
$\Sigma_b^{+}$		&$1/2^+$&$5810.56$	
&$5.0$ 		&$\Sigma_b^+ \rightarrow \Lambda_b\pi^+$ \\ 
$\Sigma_b^{0}$		&$1/2^+$&$\cdots$	
&$\cdots$	&$\cdots$\\ 
$\Sigma_b^{-}$			&$1/2^+$&$5815.64$	
&$5.3$ 		&$\Sigma_b^- \rightarrow \Lambda_b\pi^-$\\ \hline
$\Sigma_b^{*+}$		&$3/2^+$&$5830.32$	
&$9.4$ 			&$\Sigma_b^{*+} \rightarrow \Lambda_b\pi^+$\\ 
$\Sigma_b^{*0}$		&$3/2^+$&$\cdots$	
&$\cdots$	&$\cdots$\\ 
$\Sigma_b^{*-}$		&$3/2^+$&$5834.74$	
&$10.4$			&$\Sigma_b^{*-} \rightarrow \Lambda_b\pi^-$ \\ \hline
\multicolumn{1}{c}{$\Xi_b^{0}$}
&\multicolumn{1}{c}{$1/2^+$}
&\multicolumn{1}{c}{$5791.9$}	
&\multicolumn{2}{c}{(No strong decay)} \\ 
\multicolumn{1}{c}{$\Xi_b^{-}$}
&\multicolumn{1}{c}{$1/2^+$}
&\multicolumn{1}{c}{$5797.0$}	
&\multicolumn{2}{c}{(No strong decay)} \\ \hline
$\Xi_b^{'0}$			&$1/2^+$&$\cdots$	
&$\cdots$		&$\cdots$ \\ 
$\Xi_b^{'-}(5935)$		&$1/2^+$&$5935.02$	
&($<0.08$) 	&$\Xi_b^{'-} \rightarrow \Xi_b^{0(-)} \pi^{-(0)}$ \\ \hline
$\Xi_b^{0}(5945)$		&$3/2^+$&$5952.3$	
&$0.90$ 		&$\Xi_b^{'*0} \rightarrow \Xi_b^{-(0)} \pi^{+(0)}$\\ 
$\Xi_b^{-}(5955)$		&$3/2^+$&$5955.33$	
&$1.65$ 		&$\Xi_b^{'*-} \rightarrow \Xi_b^{0(-)} \pi^{-(0)}$\\ \hline \hline
  \end{tabular}
  \label{PDGdata}
\end{table}

\subsection{Theoretical framework} 
\label{sec:DecayFormula}

First, we focus on $\Sigma_Q\to \Lambda_Q\pi$ and $\Xi_Q'\to\Xi_Q\pi$ decays ($Q=c,b$).
The $\Lambda_Q$ and $\Xi_Q$ are composed of one heavy quark and the S diquark, while $\Sigma_Q$ and $\Xi_Q'$ includes the A diquark. Thus, in order to study $\Sigma_Q\to \Lambda_Q\pi$ and $\Xi_Q'\to\Xi_Q \pi$ decays, we need to focus on the couplings between the S diquark, the A diquark, and the pion in Lagrangian~(\ref{LSVExplicit}). The interaction Lagrangian is given by
\begin{equation}
\begin{split}
&\lag_{A\rightarrow S\pi}
=-\sqrt{2}iG_1\left[
A^\mu_{\{uu\}}(\partial_\mu\pi^-)S^\dag_{[ud]}\right.  \\
&\left.\hspace{1.2cm}
-A^\mu_{\{ud\}}(\partial_\mu\pi^0)S^\dag_{[ud]}
-A^\mu_{\{dd\}}(\partial_\mu\pi^+)S^\dag_{[ud]}
\right] \\
&\hspace{0.4cm}
-iG_2\left[
A^\mu_{\{ds\}}(\partial_\mu\pi^+)S^\dag_{[su]}
-A^\mu_{\{us\}}(\partial_\mu\pi^-)S^\dag_{[ds]}
\right] \\
&\hspace{0.4cm}
+i\frac{G_2}{\sqrt{2}}\left[
A^\mu_{\{us\}}(\partial_\mu\pi^0)S^\dag_{[su]}
+A^\mu_{\{ds\}}(\partial_\mu\pi^0)S^\dag_{[ds]}
\right].
\label{lagASpipm0}
\end{split}
\end{equation}
In this Lagrangian, we have rewritten the numbers in the subscripts of diquarks as the flavors: the original $A_{ij}^\mu$ and $S_{i}^\dag$ denote
\begin{eqnarray}
A_{ij}^\mu=\frac{1}{\sqrt{2}}
\begin{pmatrix}
\sqrt{2}A_{\{uu\}}^\mu	&A_{\{ud\}}^\mu		&A_{\{us\}}^\mu\\
A_{\{ud\}}^\mu			&\sqrt{2}A_{\{dd\}}^\mu	&A_{\{ds\}}^\mu\\
A_{\{us\}}^\mu			&A_{\{ds\}}^\mu			&\sqrt{2}A_{\{ss\}}^\mu
\end{pmatrix}
,
\label{flavorAdiquark}
\end{eqnarray}
\begin{eqnarray}
 S_{i}^\dag=
\begin{pmatrix}
S_{[ds]}^\dag&
S_{[su]}^\dag&
S_{[ud]}^\dag
\end{pmatrix}
,
\label{flavorSdiquark}
\end{eqnarray}
where $\{\cdots\}$ and $[\cdots]$ stand for the symmetric and anti-symmetric ordering of quarks, respectively. The new coupling constants $G_{1}$ and $G_{2}$ are given as
\begin{eqnarray}
G_1=g_1+\alpha g_2,~G_2=g_1+g_2.
\label{G1G2}
\end{eqnarray} 
From Eq.~(\ref{G1G2}), it is obvious that the difference between $G_1$ and $G_2$ comes from the violation of $SU(3)$ flavor symmetry characterized by $\alpha\neq1$.

\begin{figure}[t]
\centering
  \includegraphics[clip,width=1.05\columnwidth]{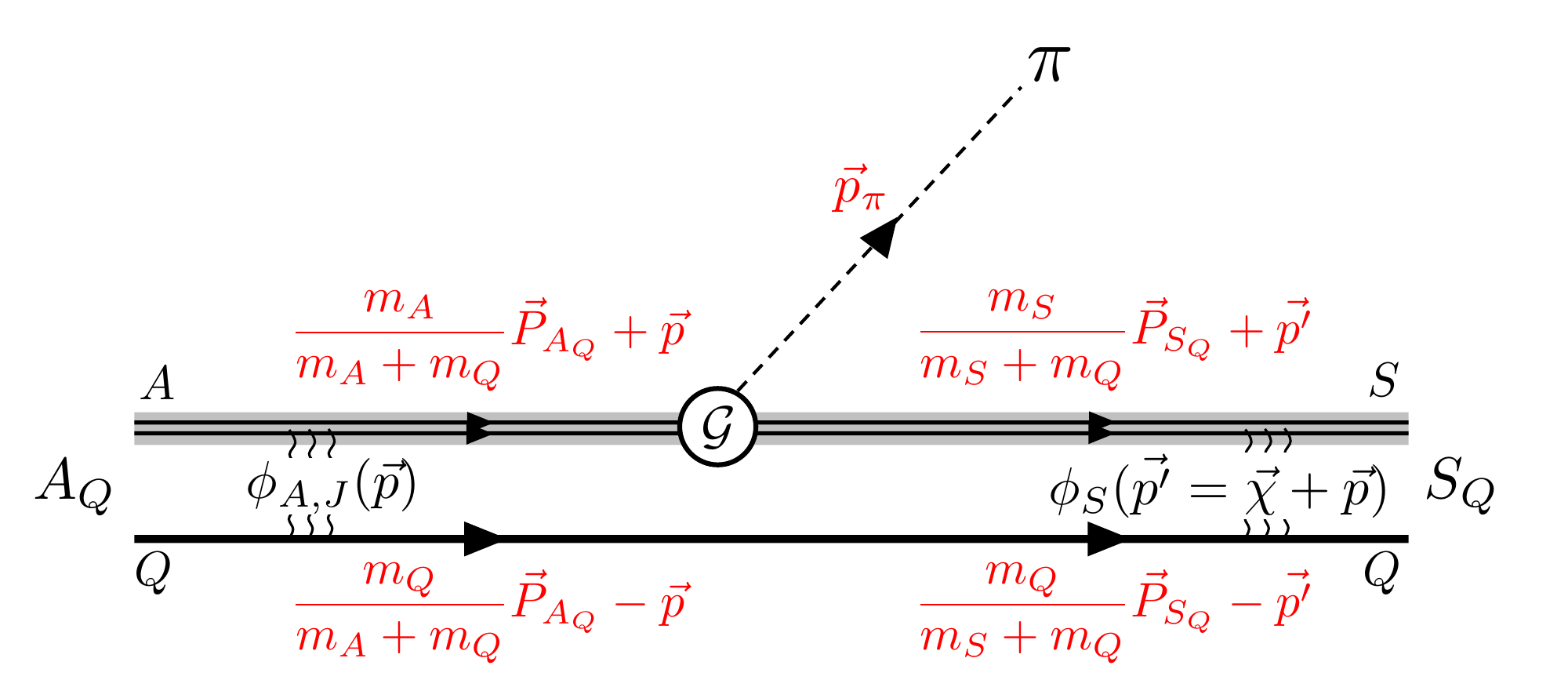}
  \caption{Kinematics employed for $A_Q\to S_Q\pi$ decays with pion momentm $\vec{p}_\pi$. $\vec{P}_{A_Q}$ and $\vec{P}_{S_Q}$ denote the momenta of initial- and final-state baryons, respectively, and $\vec{p}$ and $\vec{p'}$ are the relative momenta between a diquark and a heavy quark in the initial and final states, respectively.
$\phi_{A, J}$ and $\phi_{S}$ are the relative wave functions.
$m_{d}$ ($d=A,S$) and $m_Q$ are the diquark mass and the heavy-quark mass, respectively, and $\mathcal{G}$ is the coupling constant.} 
  \label{MD}
\end{figure}

From now on, in order to study the decays of singly heavy baryons, we employ a diquark--heavy-quark picture: While the dynamics of pion decays is determined by the Lagrangian~(\ref{lagASpipm0}) for diquarks, the property as the bound state composed of a diquark and a heavy quark is given by a quark model calculation.
In this picture, the heavy-baryon state $|d_Q\rangle$ with the energy $E_{d_Q}$ is represented as the product of the diquark state $|d\rangle$ with $E_{d}$ and a heavy-quark state $|Q\rangle$ with $E_{Q}$, superposed by
the relative wave function $\phi_d(\vec{p})$ between them. It is explicitly given in the momentum space as
\begin{equation}
\begin{split}
| d_Q (P_{d_Q},J_{d_Q})\rangle 
&= \sqrt{2E_{d_Q}}\int \frac{d^3p}{(2\pi)^3} \phi_{d}(\vec{p}) \\
&\times\frac{1}{\sqrt{2E_d }}|d(p_d, s_d)\rangle \frac{1}{\sqrt{2E_Q}}|Q(p_Q, s_Q)\rangle,
\label{baryonket}
\end{split}
\end{equation}
where the integral is over the relative momentum, $\vec{p}$, keeping
the total momentum conserved as $\vec{p}_d+\vec{p}_Q=\vec{P}_{d_Q}$. 
Here, $\vec{P}_{d_Q}$, $\vec{p}_{d}(\vec{p})\equiv\displaystyle\frac{m_d}{m_d+m_Q}\vec{P}_{d_Q}+ \vec{p}$, and 
$\vec{p}_Q(\vec{p})\displaystyle\equiv \frac{m_Q}{m_d+m_Q}\vec{P}_{d_Q}- \vec{p}$ 
(with $m_{d}$ and $m_Q$ as the diquark and heavy-quark masses, respectively)
are the momenta of the heavy baryon, diquark, and heavy quark, respectively. 
(See Fig.~\ref{MD} for the definitions.)
The energies of diquark and heavy quark also depend on the relative momentum as $E_{d}=\sqrt{|\vec{p}_{d}(\vec{p})|^2+m_{d}^2}$ and $E_{Q}=\sqrt{|\vec{p}_{Q}(\vec{p})|^2+m_{Q}^2}$.

In Eq.~(\ref{baryonket}), $J_{d_Q}$, $s_d$, and $s_Q$ represent the spins of the corresponding states.
The conservation law of spins leads to 
\begin{eqnarray}
(J_{d_Q},s_d,s_Q)=\left(\frac{1}{2},0,\frac{1}{2} \right) \label{SpinHQSS}
\end{eqnarray}
for the S diquark ($d=S$) and
\begin{eqnarray}
(J_{d_Q},s_d,s_Q)=\left(\frac{1}{2},1,\frac{1}{2}\right)\, {\rm or}\, \left(\frac{3}{2},1,\frac{1}{2}\right) \label{SpinHQSD}
\end{eqnarray}
for the A diquark ($d=A$), if there is no orbital excitation.

Note that the normalizations of the state kets are defined in the relativistic way in Eq.~(\ref{baryonket}), such as
\begin{eqnarray}
&& \langle d_Q'(\vec{P}'_{d_Q}, J'_{d_Q})|d_Q(\vec{P}_{d_Q},J_{d_Q})\rangle \nonumber\\
&& =2E_{d_Q}(2\pi)^3\delta^{(3)}(\vec{P}'_{d_Q} - \vec{P}_{d_Q})\delta_{J'_{d_Q}J_{d_Q}}, 
\label{heavyquarkpart}
\end{eqnarray}
with $E_{d_Q}=\sqrt{|\vec{P}_{d_Q}|^2+m_{d_Q}^2}$.
Similar normalizations are also applied to the state kets of the diquark and the heavy quark.
Hence, we need to include the factors of $\sqrt{2E_{d_Q}}$, $\sqrt{2E_{d}}$, and $\sqrt{2E_{Q}}$.
The normalization of the relative wave function is given by
\begin{equation}
\int\frac{d^3p}{(2\pi)^3} |\phi_d(\vec{p})|^2=1.
\end{equation}
For the pionic decays where the heavy quark is treated as a spectator,
we only need to take into account the overlaps of the diquark states $|d\rangle$ (and pion states $|\pi\rangle$) together with the wave function $\phi_d(\vec{p})$. Meanwhile, the heavy-quark part simply leads to their momentum conservation law.

Keeping the above in mind, we focus on decays of $A_Q\rightarrow S_Q\pi$ where $A_Q= \Sigma_Q$ or $\Xi'_Q$ and $S_Q=\Lambda_Q$ or $\Xi_Q$.
Using the interaction between diquarks and pions written as Eq.~(\ref{lagASpipm0}), 
the decay amplitude is given by
\begin{equation}
\int d^4x \langle S_Q (P_{S_Q}, J_{S_Q})\pi(p_\pi)|i \lag_{A\to S\pi}(x)| A_Q (P_{A_Q}, J_{A_Q})\rangle .
\label{lag_amp}
\end{equation}
From Eq.~(\ref{lagASpipm0}), the Lagrangian $\lag_{A\to S\pi}$ are expressed as the sum of the terms with operators $A^\mu_{\{qq\}}(\partial_\mu \pi) S^\dag_{[qq]}$. 
Substituting Eq.~(\ref{baryonket}) into Eq. (\ref{lag_amp}), they are calculated for each term as  
\begin{widetext}
\begin{equation}
\begin{split}
&\int d^4x \langle S_Q (P_{S_Q}, J_{S_Q})\pi(p_\pi)|i \mathcal{G}A^\mu_{\{qq\}}(x)(\partial_\mu \pi(x)) S^\dag_{[qq]}(x)| A_Q (P_{A_Q}, J_{A_Q})\rangle\\
&\hspace{3.0cm} 
=-\frac{\mathcal{G}\sqrt{E_{A_Q}E_{S_Q}}}{3}\int \frac{d^3p}{(2\pi)^3}
\frac{(p_\pi)_\mu \epsilon_A^\mu (p_A, s_A)}{\sqrt{E_AE_S}}
\phi_{S}^*(\vec{\chi} + \vec{p}) \phi_{A,J}(\vec{p})
\times (2\pi)^4 \delta^{(4)}(p_S + p_\pi - p_A).
\label{lag_amp2}
\end{split}
\end{equation}
\end{widetext}
Here the factor $\mathcal{G}$ denotes the coupling constant, in which its absolute value is given for each decay process as
\begin{equation}
|\mathcal{G}|=\left\{
\begin{array}{ll}
\sqrt{2}G_1	&(\Sigma_Q\rightarrow\Lambda_Q\pi^{\pm, 0})\\
G_2			&( \Xi'_Q\rightarrow\Xi_Q\pi^\pm)\\
G_2/\sqrt{2}	&( \Xi'_Q\rightarrow\Xi_Q\pi^{0})
\end{array}
\right..
\label{mathG}
\end{equation}
Also, $(p_\pi)_\mu$ and $\epsilon^\mu_A(p_A, s_A)$ denote the pion momentum and the polarization vector of the A diquark, respectively. 
$\phi_S(\vec{p})$ and $\phi_{A, J}(\vec{p})$ show the $S$-wave relative wave functions between the diquark and the heavy quark for $S_Q$ and $A_Q$ baryons, respectively. As shown in Eqs.~(\ref{SpinHQSS}) and~(\ref{SpinHQSD}), while $S_Q$ baryon belongs to heavy-quark spin singlet, $A_Q$ belongs to the doublet, so that the total spin of $A_Q$ is labeled as the subscript $J=1/2, 3/2$ in $\phi_{A, J}$.
The delta function in Eq.~(\ref{lag_amp2}) represents the momentum and energy conservation laws of diquarks and an emitted pion. In addition, the momenta carried by the heavy quarks preserve during the interaction. These conservation laws determine the recoil momentum $\vec{\chi} \equiv \vec{p'}-\vec{p}$ in the relative motion $\phi_S(\vec{p'})$. In particular, at the rest frame of the $A_Q$ baryon ($\vec{P}_{A_Q}=0$), $\vec{\chi}$ is written as
\begin{equation}
 \vec{\chi}= -\frac{m_Q}{m_S+m_Q}\vec{p}_\pi.
 \label{restframe_chi}
 \end{equation}
 In this frame, one can easily confirm $\vec{P}_{S_Q}=-\vec{p}_\pi$, and the momentum of A diquark coincides with the relative momentum as $\vec{p}_A = \vec{p}$. The kinematics employed in the present analysis is depicted in Fig.~\ref{MD}.


For the amplitude in Eq.~(\ref{lag_amp2}), we use an approximation for the momentum of A diquark, $\vec{p}_A$. 
That is, in our present analysis, $\vec{p}_A$ is replaced by the expectation value:
\begin{equation}
\begin{split}
&\vec{p}_A \to \langle \vec{p} \rangle \equiv \int\frac{d^3p}{(2\pi)^3} \vec{p}\, |\phi_{A,J}(\vec{p})|^2 = 0 , \\
&|\vec{p}_A|^2 \to \langle p^2\rangle \equiv \int\frac{d^3p}{(2\pi)^3} |\vec{p}|^2\,  |\phi_{A,J}(\vec{p})|^2 .
\label{Apsquare}
\end{split}
\end{equation}
In this way, the momentum squared of the S diquark is evaluated as $\langle p^2\rangle + |\vec{p}_\pi|^2$ from the momentum conservation. Then, by defining a velocity $\upsilon^2_A$ using $ \langle p^2\rangle \equiv m_A^2 \upsilon_A^2$, the energies of the A and S diquarks are written as 
\begin{eqnarray}
&&E_A= m_A\sqrt{1+\upsilon_A^2}, \\
&&E_S = m_S\sqrt{1+\frac{m_A^2}{m_S^2}\upsilon^2_A + \frac{|\vec{p}_\pi|^2}{m_S^2}},
\end{eqnarray}
respectively. This approximation is also applied to $p_A$ in the sum of the polarization vector $\epsilon^\mu_A(p_A, s_A)$ as
\begin{eqnarray}
\sum_{s_A=-1,0,1} \epsilon_A^{*\mu}(p_A, s_A)\epsilon_A^\nu(p_A, s_A)=-g^{\mu\nu}+\frac{p_A^\mu p_A^\nu}{m_A^2}.
\label{proca}
\end{eqnarray}


As a result, we find the decay amplitude $\mathcal{M}$ as 
\begin{equation}
\begin{split}
|\mathcal{M}|
=\frac{|\mathcal{G}|}{3}
\frac{\sqrt{E_{A_Q}E_{S_Q}}}{\sqrt{E_A E_S}}
\left|(p_\pi)_\mu \epsilon_A^\mu(p_A, s_A)\right|
\langle \phi_S|\phi_{A, J}\rangle,
\label{TA}
\end{split}
\end{equation}
where the inner product $\langle \phi_S|\phi_{A, J}\rangle$ is defined in the momentum space as
\begin{equation}
\langle\phi_{S}|\phi_{A, J} \rangle
\equiv \int \frac{d^3{p}}{(2\pi)^3} \phi_{S}^* ({\vec{\chi}} + \vec{p}) \phi_{A, J} (\vec{p}).
 \label{naiseki}
 \end{equation}
Finally, from the square of the amplitude~(\ref{TA}), we obtain the decay width:
\begin{equation}
\begin{split}
\Gamma[A_Q \rightarrow S_Q \pi] 
&= |\langle\phi_{S}|\phi_{A, J} \rangle|^2
\frac{|\mathcal{G}|^2 |\vec{p}_\pi| E_{S_Q}}{216\pi m_{A_Q}}  \\
&\times \left\{ \frac{|\vec{p}_{\pi}|^2(3+\upsilon_A^2) + 3E_\pi^2\upsilon_A^2}{3E_A E_S}\right\}
,
\label{diquarkwavewidth}
\end{split}
\end{equation}
where $|\vec{p}_\pi|$ is given as (with the masses of the baryons and pion)
\begin{equation}
\begin{split}
&|\vec{p}_\pi|=\frac{\sqrt{(m_{A_Q}^2-M_+^2)(m_{A_Q}^2-M_-^2)}}{2m_{A_Q}}, \\
&M_{\pm}=|m_{S_Q}\pm m_\pi |.
\label{ppi}
\end{split}
\end{equation}

The relative wave functions $\phi_S(\vec{p})$ and $\phi_{A,J}(\vec{p})$ are obtained 
by solving the Schr$\ddot{\rm o}$dinger equation $H\phi_d=E\phi_d$ 
describing a bound state of the heavy quark and diquark.
The non-relativistic diquark--heavy-quark Hamiltonian is 
\begin{equation}
H=\frac{|\vec{p}|^2}{2\mu}+m_Q+m_d+V, 
\label{ham}
\end{equation}
where the relative momentum $\vec{p}$ and the reduced mass $\mu$ are defined as
\begin{eqnarray}
&& \vec{p}=\frac{m_Q\vec{p}_d-m_d\vec{p}_Q}{m_Q+m_d},\\
&& \mu=\frac{m_Q m_d}{m_Q+m_d}, \label{reducedmass}
\end{eqnarray}
respectively.
For the potential between the heavy quark and the diquark, we use the potential in Ref.~\cite{Vdiquark} which is called the Y-potential:
\begin{equation}
V(r) = -\frac{\alpha_{\rm coul}}{r}+\lambda r+C
+({\bm s}_d \cdot {\bm s}_Q)\frac{\kappa_Q}{m_d m_Q} \frac{\Lambda^2}{r} e^{-\Lambda r},
\label{Ypot}
\end{equation}
where we include the Coulomb term with the coefficient $\alpha_{\rm coul}$, the linear confinement term with $\lambda$, the constant shift term $C$, and the spin-spin potential term with $\kappa_Q$ and a cutoff parameter $\Lambda$.
These model parameters are summarized in Table \ref{Ypotparameters} together with the masses of diquarks and heavy quarks. In Eq.~(\ref{Ypot}), we neglect other terms such as the spin-orbit potential term and the tensor term because the wave functions $\phi_S(\vec{p})$ and $\phi_{A, J}(\vec{p})$ are in the $S$-wave states.

By using the Gaussian expansion method~\cite{GEM1,GEM2}, we solve the Schr$\ddot{\rm o}$dinger equation and obtain the wave functions $\phi_S$ and $\phi_{A,J}$.

\begin{table}[tb]
  \centering
    \caption{Parameters of the Y-potential model and masses of diquarks and heavy quarks~\cite{Vdiquark}. $\mu$ in the parameter $\alpha_{\rm coul}$ is the reduced mass of diquark--heavy-quark two-body system given in Eq.~(\ref{reducedmass}). The parameters $\kappa_Q$ are determined separately for charm ($Q=c$) and bottom ($Q=b$) quarks. 
    } 
  \begin{tabular}{ l r   |   l r   } \hline\hline 
\multicolumn{2}{c|}{Y-pot. parameters}
&\multicolumn{2}{c}{Masses (MeV)}  \\ \hline 
  
  \hline
  $\alpha_{\rm coul}$		 & 60(MeV)/$\mu$	&$m_{[ud]}$	&725     \\ 
  $\lambda ({\rm GeV}^2)$ & $0.165$	&$m_{[ds/su]}$		&942       \\
  $ C_c (\rm GeV)$            & $-0.831$	&$m_{\{uu/ud/dd\}}$	&973  \\
  $ C_b (\rm GeV)$            & $-0.819$	&$m_{\{us/ds\}}$	&1116	  \\
  $\Lambda (\rm GeV)$     & $0.691$		&$m_c $			&1750   \\
  $\kappa_c$                     & $0.8586$	&$m_b $			&5112 \\
  $\kappa_b$                     & $0.6635$	&& \\ \hline\hline
  \end{tabular}
  \label{Ypotparameters}
\end{table}

\subsection{Determination of coupling constants}
\label{sec:Coupling}

By using the observed decay widths of singly heavy baryons,
we determine  the coupling constant between the diquarks and the pions in the interaction Lagrangian~(\ref{lagASpipm0})
via the formula~(\ref{diquarkwavewidth}) and Eq.~(\ref{mathG}).
More concretely, $G_1$ and $G_2$ are determined as follows:
\begin{itemize}
\item[(i)] The value of $G_1$ can be determined by the experimental data of the $\pi^\pm$ emission decays of $\Sigma_Q^{(*)}$ baryons.
On the other hand, the $\pi^0$ emission decays only give the upper limit of the decay width.
\item[(ii)] The value of $G_2$ is determined by the experimental data of the decays of $\Xi^{'}_Q$ baryons with $J= 3/2$.
This is because there are no available data of $\Xi_Q'$ with $J=1/2$ in PDG.
\end{itemize}


\begin{table}[tb]
  \centering
    \caption{
The coupling constant $G_1$ determined from the one-pion decay widths of $\Sigma_Q$ baryons. The experimental data of decay widths are also shown in units of MeV.
  }
  \begin{tabular}{ c  c | c  c  | c } \hline\hline
       \multicolumn{1}{c }{Baryon}
              &\multicolumn{1}{c |}{$J^P$}
       &\multicolumn{1}{c}{Decay mode}   
         &\multicolumn{1}{c |}{Decay width (MeV)}   
       &\multicolumn{1}{c}{$G_1$} 
        \\ \hline
       
       \hline
$\Sigma_c^{++}(2455)$&$1/2^+$	&$\Sigma_c^{++} \rightarrow \Lambda_c \pi^+$		&1.89  	&21.16 \\ 
$\Sigma_c^{0}(2455) $&$1/2^+$	&$\Sigma_c^{0} \rightarrow \Lambda_c \pi^-$		&1.83  	&20.92\\ \hline
$\Sigma_c^{++}(2520) $&$3/2^+$	&$\Sigma_c^{*++} \rightarrow \Lambda_c \pi^+$	&14.78	&25.93\\ 
$\Sigma_c^{0}(2520) $&$3/2^+$	&$\Sigma_c^{*0} \rightarrow \Lambda_c \pi^-$		&15.3  	&26.37\\ \hline

\hline
$\Sigma_b^{+} $&$1/2^+$	&$\Sigma_b^+ \rightarrow \Lambda_b\pi^+$	&5.0  	&21.93 \\ 
$\Sigma_b^{-} $&$1/2^+$	&$\Sigma_b^- \rightarrow \Lambda_b\pi^-$	&5.3  	&21.12\\ \hline
$\Sigma_b^{*+} $&$3/2^+$	&$\Sigma_b^{*+} \rightarrow \Lambda_b\pi^+$&9.4  	&23.77\\ 
$\Sigma_b^{*-} $&$3/2^+$	&$\Sigma_b^{*-} \rightarrow \Lambda_b\pi^-$	&10.4	&23.88\\ \hline\hline
  \end{tabular}
  \label{G1parameter}
\end{table}

\begin{table}[tb]
  \centering
    \caption{
 The coupling constant $G_2$ determined from the one-pion decay widths of $\Xi_Q^{'}$ baryons. The experimental data of decay widths are also shown in units of MeV. Note that the values of partial decay widths with the asterisk (*) are the predicted widths in our present work~\cite{Vdiquark}.
  }
  \begin{tabular}{ c  c  |c c | c} \hline\hline
       \multicolumn{1}{c}{Baryon}
       &\multicolumn{1}{c|}{$J^P$}
       &\multicolumn{1}{c}{Decay mode}   
         &\multicolumn{1}{c|}{Decay width (MeV)}   
                  &\multicolumn{1}{c}{$G_2$} 
     
        \\ \hline
       
       \hline

				&		&$\Xi_c^{'*+} \rightarrow \Xi_c \pi$			&2.14&\\ 
\cline{3-4}					
$\Xi_c^{+}(2645)$	&$3/2^+$	&$\Xi_c^{'*+} \rightarrow \Xi_c^0 \pi^+$		&1.32*&26.73\\ 

				&		&$\Xi_c^{'*+} \rightarrow \Xi_c^+ \pi^0$		&0.82*&\\
\hline

				&		&$\Xi_c^{'*0} \rightarrow \Xi_c \pi$			&2.35&\\ 
\cline{3-4}					
$\Xi_c^{0}(2645)$	&$3/2^+$	&$\Xi_c^{'*0} \rightarrow \Xi_c^+ \pi^-$		&1.56*&27.07 \\ 

				&		&$\Xi_c^{'*0} \rightarrow \Xi_c^0 \pi^0$		&0.79*&\\
\hline

\hline

				&		&$\Xi_b^{'*0} \rightarrow \Xi_b \pi$			&0.90&\\
\cline{3-4} 
$\Xi_b^{0}(5945)$	&$3/2^+$	&$\Xi_b^{'*0} \rightarrow \Xi_b^- \pi^+$		&0.50*&24.76\\ 

				&		&$\Xi_b^{'*0} \rightarrow \Xi_b^0 \pi^0$		&0.41*&\\
 \hline

				&		&$\Xi_b^{'*-} \rightarrow \Xi_b \pi$			&1.65&\\ 
\cline{3-4}
$\Xi_b^{-}(5955)$	&$3/2^+$	&$\Xi_b^{'*-} \rightarrow \Xi_b^0 \pi^-$		&1.13*&29.55\\ 

				&		&$\Xi_b^{'*-} \rightarrow \Xi_b^- \pi^0$		&0.52*&\\
\hline\hline
  \end{tabular}
  \label{G2parameter}
\end{table}

In Tables \ref{G1parameter} and \ref{G2parameter}, we show the obtained $G_1$ and $G_2$ for each decay.
From these tables, one sees that both the values of $G_1$ and $G_2$ range from 20 to 30, where $21 < G_1 < 27$ and $24 < G_2 < 30$, so that $G_2$ tends to be a bit larger than $G_1$. We note that, depending on used experimental data, both $G_1$ and $G_2$ have the indeterminacy of approximately 6.

We comment on the errors of experimental data. The values of $G_1$ and $G_2$ shown in Tables \ref{G1parameter} and \ref{G2parameter} are evaluated from the averages of the experimental data. If we take into account the errors of these data, we can also estimate the errors of $G_1$ and $G_2$. From all of the eight results of $G_1$ and the four results of $G_2$, we have confirmed the margin of errors of $G_1$ and $G_2$ are from 2 to 7.
For example, according to PDG \cite{PDG}, the decay width and baryon masses of $\Sigma_c^{++}\rightarrow \Lambda_c \pi^+$ strong decay are $\Gamma[\Sigma_c^{++}\rightarrow \Lambda_c \pi^+]=1.89^{+0.09}_{-0.18}$ MeV, $m_{\Sigma_c^{++}}=2453.97\pm 0.14$ MeV, and $m_{\Lambda_c}=2286.46\pm 0.14$ MeV, respectively. From analysis using the upper and lower errors of these values, the errors of coupling constant $G_1$ are evaluated as $G_1=21.16^{+0.63}_{-1.15}$. Similarly, for the $\Xi_b^{'*-} \rightarrow \Xi_b \pi$ decay, the errors of its decay width and baryon masses lead to $G_2=29.55^{+3.38}_{-3.55}$. In the following analysis, for simplicity, we neglect these errors.

We take the isospin and spin averages of $G_1$'s in Table \ref{G1parameter}, and obtain
\begin{eqnarray}
&&\bar{G}_1^{c(1/2)}\equiv
\frac{1}{2}\left(G_1|_{\Sigma_c^{++}(2455)}+G_1|_{\Sigma_c^0(2455)}\right) \simeq 21.04~,\nonumber\\
&&\bar{G}_1^{c(3/2)}\equiv
\frac{1}{2}\left(G_1|_{\Sigma_c^{++}(2520)}+G_1|_{\Sigma_c^{0}(2520)}\right)\simeq 26.15~,\nonumber\\
&&\bar{G}_1^{b(1/2)}\equiv
\frac{1}{2}\left(G_1|_{\Sigma_b^{+}}+G_1|_{\Sigma_b^{-}}\right)\simeq 21.53~,\nonumber\\
&&\bar{G}_1^{b(3/2)}\equiv
\frac{1}{2}\left(G_1|_{\Sigma_b^{*+}}+G_1|_{\Sigma_b^{*-}}\right)\simeq 23.82~,
\label{G1isospinave}
\end{eqnarray}
and
\begin{equation}
\begin{split}
&G_1^{c}\equiv
\frac{1}{3}\left(\bar{G}_1^{c(1/2)} + 2\bar{G}_1^{c(3/2)} \right)\simeq 24.45~,\\
&G_1^{b}\equiv
\frac{1}{3}\left(\bar{G}_1^{b(1/2)} + 2\bar{G}_1^{b(3/2)} \right)\simeq 23.06~.
\label{G1ave}
\end{split}
\end{equation}
For $G_2$'s in Table \ref{G2parameter}, by taking only the isospin average, we obtain
\begin{eqnarray}
\begin{split}
&&G_2^{c}\equiv
\frac{1}{2}\left(G_2|_{\Xi_c^{+}(2645)}+G_2|_{\Xi_c^0(2645)}\right)\simeq 26.90~,\\
&&G_2^{b}\equiv
\frac{1}{2}\left(G_2|_{\Xi_b^{0}(5945)}+G_2|_{\Xi_b^{-}(5955)}\right)\simeq 27.16~.
\label{G2ave}
\end{split}
\end{eqnarray}
The values of $G_1$ and $G_2$ in Eqs.~(\ref{G1ave}) and~(\ref{G2ave}) indicate that the violation of $SU(3)$ flavor symmetry characterized by the difference between $G_1$ and $G_2$ are considerably small.
In addition, the violation of heavy-quark flavor symmetry characterized by the difference between $G_{1,2}^c$ and $G_{1,2}^b$ is also small.

Using Eq. (\ref{G1G2}), we determine the $g_1$ and $g_2$ parameters in the original effective Lagrangian~(\ref{SVall}).
Since the $g_2$ term breaks the $U(1)_A$ symmetry while the $g_1$ term does not, the influence of the $U(1)_A$ anomaly on the diquarks are quantified by translating the values of $G_1$ and $G_2$ into those of $g_1$ and $g_2$. When we take $g_s=3$ for the quark-meson coupling constant, the dimensionless quantity $\alpha$ in Eq.~(\ref{effectivequarkmass}) is estimated to be $\alpha\simeq 1.73$. Thus, from the relations in Eq.~(\ref{G1G2}), the magnitudes of $g_1$ and $g_2$ are estimated as
\begin{eqnarray}
&&(g^c_1, g^c_2)=(30.26,  -3.36    ),\hspace{0.8cm}
\label{g1g2CD}\\
&&(g^b_1, g^b_2)=(32.78,  -5.62    ),\hspace{0.8cm}
\label{g1g2BD}
\end{eqnarray}
from Eqs.~(\ref{G1ave}) and (\ref{G2ave}) for charmed and bottom sectors, respectively. From these values, one immediately sees that the magnitude of $g_1$ is much larger than $g_2$: The ratio of $g_2$ to $g_1$ is evaluated to be $|g^c_2/g^c_1|\simeq 0.11$ for the charm sector and $|g^b_2/g^b_1|\simeq 0.17$ for the bottom one. Therefore, we can conclude that the $U(1)_A$ anomaly effects to interactions between the diquarks and pions are small.
Note that the same conclusion is obtained from a chiral effective model for singly heavy baryons~\cite{Kawakami:2018olq, Kawakami:2019hpp, Suenaga:2022ajn}.
(In Appendix \ref{Sec_App}, we compare the results from our diquark model and from the model for singly heavy baryons.)


Next, in Tables~\ref{Sigpi0decay} and \ref{Xib12decay}, we show our predictions for decay widths unknown in PDG, or Table \ref{PDGdata}. First, we tabulate the predictions of $\pi^0$ emission decays of $\Sigma_Q$ baryons in Table~\ref{Sigpi0decay}. In this evaluation, we have used the isospin average coupling $G_1$ in Eq.~(\ref{G1isospinave}). For the unknown masses of $\Sigma_b$ baryons, we have used the values predicted by the diquark--heavy-quark model in Ref.~\cite{Vdiquark}.
From this table, the predicted widths of $\Sigma_c^{+} (1/2^+)$ and $\Sigma_c^{*+} (3/2^+)$ are below the experimental upper limits. 
In this sense, our predictions are consistent with the experiments. 
Also, in Table \ref{Xib12decay}, we show our predictions of one pion decay widths of $\Xi'_b$ baryons with $J^P=1/2^+$. 
Since there is no experimental value for the mass of $\Xi_b'^0 (1/2^+)$, again we have adopted that from Ref.~\cite{Vdiquark}. 
Using the isospin averaged value of coupling $G_2=27.16$ for $\Xi_b (3/2^+)$ 
, however, the predicted decay widths are quite different from the upper limit of the decay width of $\Xi_b^{'-} \rightarrow \Xi_b \pi$: $\Gamma_{\rm max}=0.08~ \rm MeV$. To satisfy this condition, the coupling constant $G_2$ needs to be less than $15.61$.

\begin{table}[tb!]
  \centering
    \caption{Our predictions of $\pi^0$ emission decay widths of $\Sigma_Q$ baryons in units of MeV. The asterisk (*)  indicates that the mass is taken from \cite{Vdiquark}. For the decay width, (cal.) and (exp.) represent our predictions and experimental data~\cite{PDG}, respectively.
  }
  \begin{tabular}{ l  c  c | c c c } \hline\hline
       \multicolumn{1}{l }{Baryon ($J^P$)}
       &\multicolumn{1}{c}{Mass}
       &\multicolumn{1}{c|}{$G_1$} 
       &\multicolumn{1}{c}{Decay mode}   
         &\multicolumn{2}{c}{Decay width}   \\
       \multicolumn{3}{c |}{}
       &\multicolumn{1}{c }{} 
       &\multicolumn{1}{c}{(cal.)}
         &\multicolumn{1}{c}{(exp.)}   \\
       \multicolumn{1}{c }{}
       &\multicolumn{1}{c }{(MeV)}
       &\multicolumn{1}{c |}{}
       &\multicolumn{1}{c }{} 
       &\multicolumn{1}{c}{(MeV)}
         &\multicolumn{1}{c}{(MeV)}   \\ \hline
       
       \hline
$\Sigma_c^{+} (1/2^+)$	&2452.9	&21.04	&$\Sigma_c^{+} \rightarrow \Lambda_c \pi^0$		&2.05&($<4.6$) \\ 
$\Sigma_c^{*+} (3/2^+)$	&2517.5	&26.15	&$\Sigma_c^{*+} \rightarrow \Lambda_c \pi^0$		&15.42&($<17$)\\ 
\hline

\hline
$\Sigma_b^{0} (1/2^+)$	&5809.88*	&21.53	&$\Sigma_b^0 \rightarrow \Lambda_b\pi^0$		&5.10&$\cdots$ \\ 
$\Sigma_b^{*0} (3/2^+)$	&5829.31*	&23.82	&$\Sigma_b^{*0} \rightarrow \Lambda_b\pi^0$		&9.75&$\cdots$\\ \hline\hline
  \end{tabular}
  \label{Sigpi0decay}
\end{table}
\begin{table}[tb!]
  \centering
    \caption{Our predictions of $\pi$ emission decay widths of $\Xi'_b (1/2^+)$ baryons in units of MeV. The asterisk (*)  indicates that the mass is taken from \cite{Vdiquark}. For the decay width, (cal.) and (exp.) represent our predictions and experimental data~\cite{PDG}, respectively.}
  \begin{tabular}{ c  c  c |c c c } \hline\hline
       \multicolumn{1}{l }{Baryon ($J^P$)}
       &\multicolumn{1}{c}{Mass}
       &\multicolumn{1}{c|}{$G_2$} 
       &\multicolumn{1}{c}{Decay mode}   
         &\multicolumn{2}{c}{Decay width}   \\
       \multicolumn{3}{c |}{}
       &\multicolumn{1}{c }{} 
       &\multicolumn{1}{c}{(cal.)}
         &\multicolumn{1}{c}{(exp.)}   \\
                \multicolumn{1}{c }{}
       &\multicolumn{1}{c }{(MeV)}
       &\multicolumn{1}{c |}{}
       &\multicolumn{1}{c }{} 
       &\multicolumn{1}{c}{(MeV)}
         &\multicolumn{1}{c}{(MeV)}   \\ \hline

        \hline
      

 
\hline

					&		&		&$\Xi_b^{'0} \rightarrow \Xi_b \pi$		&0.14&$\cdots$\\ 
\cline{4-6}
       \multicolumn{1}{l }{$\Xi_b'^{0}(1/2^+)$	}
       &\multicolumn{1}{c}{5934.16*}
       &\multicolumn{1}{c|}{27.16	} 
       &\multicolumn{1}{c}{$\Xi_b^{'0} \rightarrow \Xi_b^- \pi^+$}   
         &\multicolumn{2}{c}{(No strong decay)}   \\
 
					&		&		&$\Xi_b^{'0} \rightarrow \Xi_b^0 \pi^0$	&0.14&$\cdots$\\
\hline

\hline

%
					&		&		&$\Xi_b^{'-} \rightarrow \Xi_b \pi$		&0.24&($<0.08$)\\ 
\cline{4-6}
$\Xi_b'^{-}(1/2^+)$		&5935.02	&27.16	&$\Xi_b^{'-} \rightarrow \Xi_b^0 \pi^-$	&0.17&$\cdots$\\
 
					&		&		&$\Xi_b^{'-} \rightarrow \Xi_b^- \pi^0$	&0.07&$\cdots$\\ 
\hline\hline
  \end{tabular}
  \label{Xib12decay}
\end{table}


\subsection{Decays under chiral symmetry restoration}
In this subsection, we discuss a chiral symmetry restoration effect for decays.
Chiral symmetry tends to be restored in an extreme environment such as high temperature and/or high baryon density. In chiral effective models, such chiral symmetry restoration can be described as a decrease of the absolute value of VEV toward zero, i.e., $\langle \Sigma \rangle \rightarrow 0$.
In order to incorporate the change of VEV, in Refs.~\cite{Vdiquark, Kim:2022mpa}, the authors introduced a parameter $x$ characterizing chiral symmetry breaking: $\langle \Sigma \rangle = x\langle\Sigma\rangle|_{\rm vac}$.
In this treatment, the range of $x$ is limited to $0\le x \le 1$.
$x=0$ corresponds to chiral symmetry restored phase while $x=1$ corresponds to the ordinary chiral-symmetry broken vacuum. Then, the mass formulas of nonstrange S and A diquarks are given as a function of $x$~\cite{Vdiquark}:
\begin{eqnarray}
&&m_{[ud]}=\sqrt{m_{S0}^2-(x+\alpha-1)m_{S1}^2-x^2m_{S2}^2}, 
\label{uddiquark}\\
&&m_{\{uu/ud/dd\}}=\sqrt{m_{V0}^2+x^2(m_{V1}^2 + 2m_{V2}^2)},
\label{nndiquark} 
\end{eqnarray}
where the diquark mass parameters are~\cite{scalar,Vdiquark}%
\footnote{In obtaining Eqs.~(\ref{udparameter}) and~(\ref{nnparameter}), we use $\alpha=5/3$ (for $g_s\simeq 3.68$), which is slightly different from the one, $\alpha\simeq 1.73$ (for $g_s=3$), used in determining Eqs.~(\ref{g1g2CD}) and (\ref{g1g2BD}). Throughout this subsection, we use $\alpha=5/3$ and the values of the coupling constants are also modified accordingly. The difference is small and does not change the results qualitatively.}%
\begin{eqnarray}
&&(m_{S0}^2, m_{S1}^2, m_{S2}^2)=((1119)^2, (690)^2, -(258)^2), 
\label{udparameter}\\
&&(m_{V0}^2, m_{V1}^2, m_{V2}^2)=((708)^2, -(757)^2, (714)^2),
\label{nnparameter}
\end{eqnarray}
in units of ${\rm MeV}^2$.
The mass formulas~(\ref{uddiquark}) and~(\ref{nndiquark}) indicates that, at the chiral restoration point $x=0$, the S diquark mass becomes heavier than the A diquark one, i.e., $m_{[ud]}>m_{\{uu/ud/dd\}}$, which is contrast to the normal mass ordering $m_{\{uu/ud/dd\}}>m_{[ud]}$ realized in the vacuum at $x=1$. In particular, within our parameters, the mass inversion occurs at $x\simeq 0.6$~\cite{Vdiquark}. 


When we take into account the chiral symmetry restoration, not only the diquark masses in Eqs.~(\ref{uddiquark}) and~(\ref{nndiquark}) but also the coupling constant are modified due to $\langle\Sigma\rangle$ in the $g_2$ term in Eq.~(\ref{SV2}). That is, now the couplings $G_1$ and $G_2$ defined in Eq.~(\ref{G1G2}) are replaced by
\begin{eqnarray}
G_1=g_1+x\alpha g_2,~G_2=g_1+xg_2.
\end{eqnarray} 
 
\begin{figure}[b]
\centering
  \includegraphics[clip,width=1.02\columnwidth]{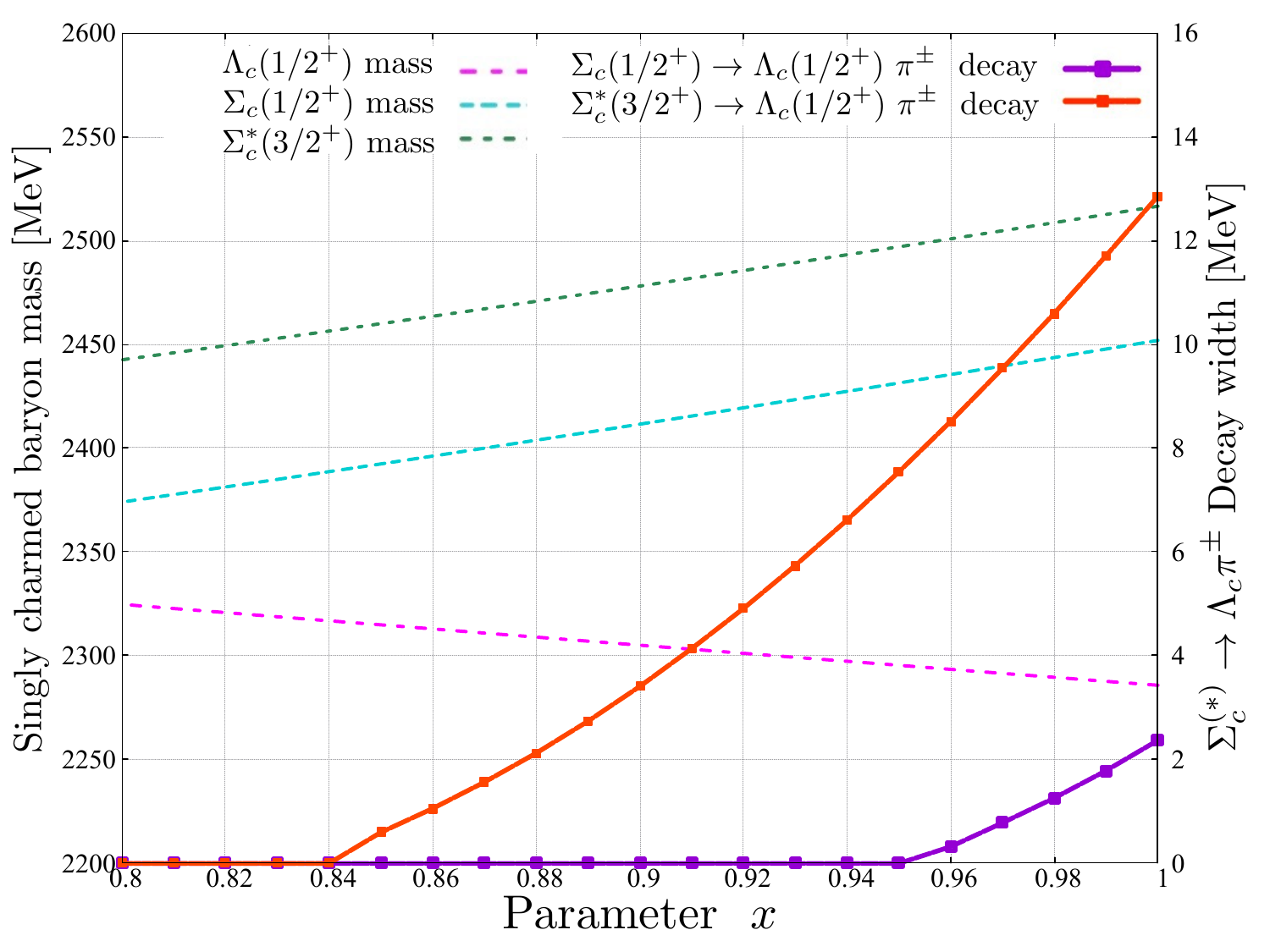}
  \caption{Dependence of the singly charmed baryon masses and decay widths of $\Sigma_c^{(*)} \rightarrow \Lambda_c \pi^\pm$ on the chiral-symmetry breaking parameter $x$ in the range of $0.80 \le x \le 1.00$.} 
  \label{figureCdecay}
\end{figure}
\begin{figure}[t]
\centering
  \includegraphics[clip,width=1.02\columnwidth]{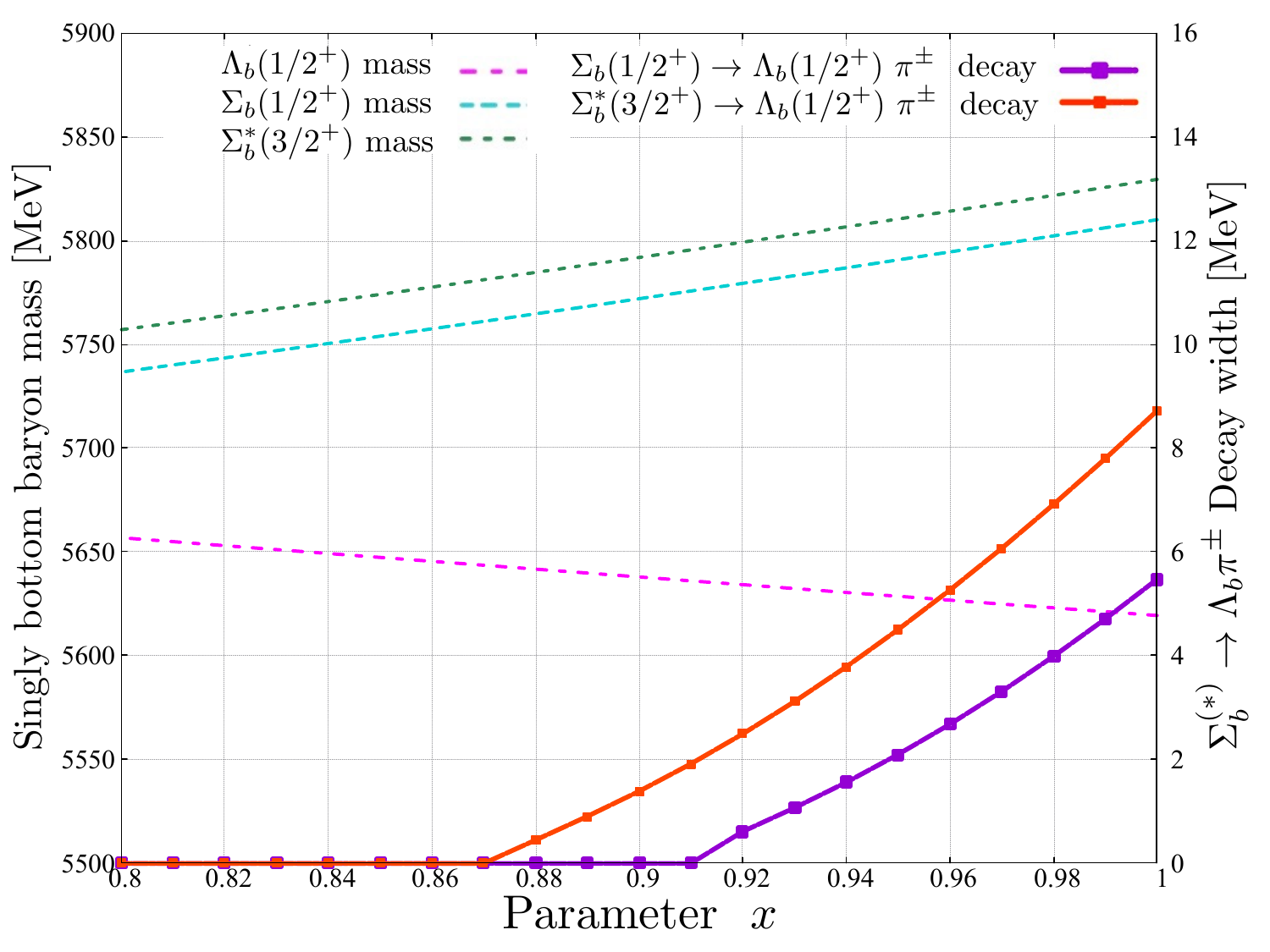}
  \caption{Dependence of the singly bottom baryon masses and decay widths of $\Sigma_b^{(*)} \rightarrow \Lambda_b \pi^\pm$ on the chiral-symmetry breaking parameter $x$ in the range of $0.80 \le x \le 1.00$.} 
  \label{figureBdecay}
\end{figure}

In Figs.~\ref{figureCdecay} and \ref{figureBdecay}, we show the $x$ dependence of the masses of $\Lambda_Q$, $\Sigma_Q$, and $\Sigma_Q^*$ and the decay widths of $\Sigma_Q^{(*)} \rightarrow \Lambda_Q \pi^{\pm}$.
We find that, as $x$ decreases, the $\Lambda_Q$ becomes heavier while $\Sigma_Q$ and $\Sigma_Q^*$ become lighter, which is naively expected from the $x$ dependence of diquark mass formulas~(\ref{uddiquark}) and~(\ref{nndiquark}).
Due to the changes of baryon masses and coupling constants, the decay widths monotonically decrease and finally vanish at a certain $x$.
In other words, a decay channel is prohibited below a threshold of $x$. Such thresholds, $x_\mathrm{th}$, are estimated to be $x_\mathrm{th} \simeq 0.95$, $0.84$, $0.91$, and $0.87$ for $\Sigma_c$, $\Sigma_c^*$, $\Sigma_b$, and $\Sigma_b^*$, respectively.



\section{Conclusion} \label{Sec_4}

In this paper, we have studied the interaction between the spin 0 and spin 1 diquarks from the viewpoints of a chiral $SU(3)_R\times SU(3)_L$ symmetry. Four diquarks with color $\bar{3}$ are considered, in which the pair of scalar (S, $0^+$) and pseudoscalar (P, $0^-$) diquarks are assigned to the chiral $(\bar 3, 1)+(1, \bar 3)$ representation, while vector (V, $1^-$) and axial-vector (A, $1^+$) diquarks are the other partners belonging to the chiral $(3, 3)$ representation. 

The chiral effective Lagrangian is constructed in the form of linear sigma model, which describes the interaction between scalar and vector diquarks with the coupling constants $g_1$ and $g_2$. The term with the coupling $g_1$ represents the interaction between a diquark and one meson, while the term with $g_2$ does the interaction between a diquark and two mesons. We have found that the $g_2$ term breaks the $U(1)_A$ symmetry, and it also leads to violation of the flavor $SU(3)$ symmetry due to explicit chiral symmetry breaking. 

These coupling constants are determined from the decay width formula (\ref{diquarkwavewidth}) and the experimental data of the one-pion emission decays of singly heavy baryons, where we have regarded these baryons as the two-body systems of a scalar or an axial-vector diquark and a heavy quark.
Our findings from our model are as follows:
\begin{itemize}
\item[(i)] Our main finding is that the magnitude of $g_2$ is much smaller than that of $g_1$, as shown in Eqs.~(\ref{g1g2CD}) and (\ref{g1g2BD}), which indicates that the effects originated from the $g_2$ term, such as the $U(1)_A$ anomaly and the flavor $SU(3)$ symmetry breaking, should be strongly suppressed.
%
\item[(ii)] We have predicted experimentally unknown decay widths, such as $\Sigma_Q^{(*)} \rightarrow \Lambda_Q \pi^0$ and $\Xi_b^{'}(1/2^+) \rightarrow \Xi_b \pi$, as summarized in Tables~\ref{Sigpi0decay} and \ref{Xib12decay}.
\item[(iii)] We have investigated the modification of masses and decay widths of $\Lambda_Q$ and $\Sigma_Q^{(*)}$ baryons by changing a parameter $x$ characterizing the magnitude of chiral symmetry breaking. As shown in Figs.~\ref{figureCdecay} and \ref{figureBdecay}, as $x$ decreases, $\Lambda_Q$ and $\Sigma_Q^{(*)}$ baryon masses become closer to each other due to the change of nonstrange S and A diquark masses~\cite{Vdiquark}. As a result, the decay widths of $\Sigma_Q^{(*)} \rightarrow \Lambda_Q \pi^{\pm}$ decrease and finally vanish below a threshold of $x$. 
\end{itemize}

In this work, we have focused on only the chiral effective model with the spin 0 and 1 diquarks with color $\bar{3}$. 
It may be interesting to improve our model by introducing other interactions or other diquark degrees of freedom.
An example is the one-pion interactions between vector and axial-vector diquarks. 
Also, the tensor diquarks with color $\bar{3}$ and flavor $6$ \cite{Shuryak:2003zi, Hong:2004xn, Shuryak:2005pk} could be important for the improvement of the diquark model, which is useful to describe the negative-parity excited states of flavor sextet singly heavy baryons ($\Sigma_Q$, $\Xi'_Q$, and $\Omega_Q$).

It is also important to examine the properties of diquarks inside hadrons under chiral symmetry restored environments such as finite temperature \cite{Sateesh:1991jt, Lee:2007wr, Oh:2009zj} and/or density. 
As predicted in this work, the suppression (and also prohibition) of $\Sigma_Q^{(*)} \rightarrow \Lambda_Q \pi^{\pm}$ decay widths would be a good signal indicating chiral symmetry restoration via diquarks.

\section*{Acknowledgments}
This work was supported by the RIKEN special postdoctoral researcher program (D.S.), and by Grants-in-Aid for Scientific Research No.~JP20K03959 (M.O.), No.~JP21H00132 (M.O.), No.~JP17K14277 (K.S.), and No.~JP20K14476 (K.S.),
and for JSPS Fellows No.~JP21J20048 (Y.K.) from Japan Society for the Promotion of Science.

\appendix

\section{Effective Lagrangian of singly heavy baryons} \label{Sec_App}
In the main text, we have constructed the chiral effective Lagrangians of the diquarks.
In this appendix, we consider the chiral effective Lagrangian for singly heavy baryons (SHBs), $A_Q(=\Sigma_Q, \Xi'_Q)$ and $S_Q(=\Lambda_Q, \Xi_Q)$, and derive their decay width formulas.  By combining a heavy quark field $Q^a$ and an A or S diquark field, SHB fields are written as
\begin{eqnarray}
S_{Q,i} \equiv Q^a S^a_{i},~~
A_{Q,ij}^\mu \equiv Q^a A_{ij}^{a,\mu} ,
\end{eqnarray}
which are explicitly
\begin{eqnarray}
A_{Q,ij}^\mu=\frac{1}{\sqrt{2}}
\begin{pmatrix}
\sqrt{2}\Sigma_Q^{I=1,\mu}	&\Sigma_Q^{I=0,\mu}		&\Xi_Q^{'I=1/2,\mu}\\
\Sigma_Q^{I=0,\mu}			&\sqrt{2}\Sigma_Q^{I=-1,\mu}	&\Xi_Q^{'I=-1/2,\mu}\\
\Xi_Q^{'I=1/2,\mu}			&\Xi_Q^{'I=-1/2,\mu}			&\sqrt{2}\Omega_Q^\mu
\end{pmatrix}
,\hspace{0.3cm}
\label{AQfield}
\end{eqnarray}
\begin{eqnarray}
S_{Q,i}=
\begin{pmatrix}
\Xi_Q^{I=-1/2}\\
\Xi_Q^{I=1/2}\\
\Lambda_Q
\end{pmatrix}
.
\label{SQfield}
\end{eqnarray}

By replacing diquark fields in Eq. (\ref{lagASpipm0}) by the corresponding SHB fields, the chiral effective Lagrangian is rewritten as
\begin{equation}
\begin{split}
&\lag_{A_Q \rightarrow S_Q\pi}
=-i\frac{\tilde{G}_1}{f_\pi}\left[
\bar{\Lambda}_Q\partial_\mu\pi^-\Sigma_Q^{I=1,\mu} \right. \\
&\left. \hspace{1.2cm}
-\bar{\Lambda}_Q\partial_\mu\pi^0\Sigma_Q^{I=0,\mu}
-\bar{\Lambda}_Q\partial_\mu\pi^+\Sigma_Q^{I=-1,\mu}
\right] \\
&\hspace{1.2cm}
-i\frac{\tilde{G}_2}{\sqrt{2}f_\pi}\left[
\bar{\Xi}_Q^{I=1/2}\partial_\mu\pi^+\Xi_Q^{'I=-1/2,\mu} \right. \\
&\left. \hspace{3.0cm}
-\bar{\Xi}_Q^{I=-1/2}\partial_\mu\pi^-\Xi_Q^{'I=1/2,\mu}
\right] \\
&\hspace{1.2cm}
+i\frac{\tilde{G}_2}{2 f_\pi}\left[
\bar{\Xi}_Q^{I=1/2}\partial_\mu\pi^0\Xi_Q^{'I=1/2,\mu} \right. \\
&\left. \hspace{3.0cm}
+\bar{\Xi}_Q^{I=-1/2}\partial_\mu\pi^0\Xi_Q^{'I=-1/2,\mu}
\right] ,
\label{lagAQSQpipm0}
\end{split}
\end{equation}
where all coefficients are divided by $\sqrt{2}f_\pi\simeq 130$ MeV to adjust these units.

According to Refs. \cite{Kawakami:2018olq, Kawakami:2019hpp, Suenaga:2022ajn}, 
the $A_Q\rightarrow S_Q \pi$ decay width formula is expressed as
\begin{eqnarray}
&&\Gamma[A_Q \rightarrow S_Q  \pi]
=\frac{|\tilde{\mathcal{G}}|^2 |\vec{p}_\pi|^3}{12\pi m_{A_Q}}(m_{S_Q}+E_{S_Q}),
\label{SHBdecaywidth}\\
&&|\tilde{\mathcal{G}}|=\left\{
\begin{array}{ll}
\tilde{G}_1/ f_\pi			&(\Sigma_Q\rightarrow\Lambda_Q\pi^{\pm, 0})\\
\tilde{G}_2/(\sqrt{2}f_\pi)	&( \Xi'_Q\rightarrow\Xi_Q\pi^\pm)\\
\tilde{G}_2/(2f_\pi)			&( \Xi'_Q\rightarrow\Xi_Q\pi^{0})
\end{array}
\right. ,
\label{SHBmathG}
\end{eqnarray}
where $|\vec{p}_\pi|$ is given in Eq.~(\ref{ppi}).
Using Eqs.~(\ref{SHBdecaywidth}) and (\ref{SHBmathG}), the parameter $\tilde{G}_1$ and $\tilde{G}_2$ are determined by substituting the baryon masses and the decay widths in Table \ref{PDGdata} with the pion mass $m_\pi$ and the pion decay constant $f_\pi$ given in this section.


\begin{table}[tb]
  \centering
    \caption{
  Numerical results of the coupling constant $\tilde{G}_1$ from the SHB model. In each $\Sigma_Q\rightarrow\Lambda_Q\pi$ decay mode, the experimental data of decay widths are also summarized in units of MeV.
  }
  \begin{tabular}{ c  c | c  c  | c } \hline\hline
       \multicolumn{1}{c }{Baryon}
              &\multicolumn{1}{c |}{$J^P$}
       &\multicolumn{1}{c}{Decay mode}   
         &\multicolumn{1}{c |}{Decay width (MeV)}   
       &\multicolumn{1}{c}{$\tilde{G}_1$} 
        \\ \hline
       
       \hline
$\Sigma_c^{++}(2455)$&$1/2^+$	&$\Sigma_c^{++} \rightarrow \Lambda_c \pi^+$		&1.89  	&0.67 \\ 
$\Sigma_c^{0}(2455) $&$1/2^+$	&$\Sigma_c^{0} \rightarrow \Lambda_c \pi^-$		&1.83  	&0.67\\ \hline
$\Sigma_c^{++}(2520) $&$3/2^+$	&$\Sigma_c^{*++} \rightarrow \Lambda_c \pi^+$	&14.78	&0.69\\ 
$\Sigma_c^{0}(2520) $&$3/2^+$	&$\Sigma_c^{*0} \rightarrow \Lambda_c \pi^-$		&15.3  	&0.70\\ \hline

\hline
$\Sigma_b^{+} $&$1/2^+$	&$\Sigma_b^+ \rightarrow \Lambda_b\pi^+$	&5.0  		&0.63 \\ 
$\Sigma_b^{-} $&$1/2^+$	&$\Sigma_b^- \rightarrow \Lambda_b\pi^-$	&5.3  		&0.60\\ \hline
$\Sigma_b^{*+} $&$3/2^+$	&$\Sigma_b^{*+} \rightarrow \Lambda_b\pi^+$&9.4  		&0.65\\ 
$\Sigma_b^{*-} $&$3/2^+$	&$\Sigma_b^{*-} \rightarrow \Lambda_b\pi^-$	&10.4	&0.65\\ \hline\hline
  \end{tabular}
  \label{SHBG1parameter}
\end{table}
\begin{table}[tb]
  \centering
    \caption{
  Numerical results of the coupling constant $\tilde{G}_2$ from the SHB model. In each $\Xi'_Q\rightarrow\Xi_Q\pi$ decay mode, the experimental data of decay widths are also summarized in units of MeV. Note that the values of partial decay widths with the asterisk (*) are the calculated values in this work.
  }
  \begin{tabular}{ c  c  |c c | c} \hline\hline
       \multicolumn{1}{c}{Baryon}
       &\multicolumn{1}{c|}{$J^P$}
       &\multicolumn{1}{c}{Decay mode}   
         &\multicolumn{1}{c|}{Decay width (MeV)}   
                  &\multicolumn{1}{c}{$\tilde{G}_2$} 
     
        \\ \hline
       
       \hline

				&		&$\Xi_c^{'*+} \rightarrow \Xi_c \pi$			&2.14&\\ 
\cline{3-4}					
$\Xi_c^{+}(2645)$	&$3/2^+$	&$\Xi_c^{'*+} \rightarrow \Xi_c^0 \pi^+$		&1.29*&0.65\\ 

				&		&$\Xi_c^{'*+} \rightarrow \Xi_c^+ \pi^0$		&0.85*&\\
\hline

				&		&$\Xi_c^{'*0} \rightarrow \Xi_c \pi$			&2.35&\\ 
\cline{3-4}					
$\Xi_c^{0}(2645)$	&$3/2^+$	&$\Xi_c^{'*0} \rightarrow \Xi_c^+ \pi^-$		&1.55*&0.65 \\ 

				&		&$\Xi_c^{'*0} \rightarrow \Xi_c^0 \pi^0$		&0.80*&\\
\hline

\hline

				&		&$\Xi_b^{'*0} \rightarrow \Xi_b \pi$			&0.90&\\
\cline{3-4} 
$\Xi_b^{0}(5945)$	&$3/2^+$	&$\Xi_b^{'*0} \rightarrow \Xi_b^- \pi^+$		&0.44*&0.69\\ 

				&		&$\Xi_b^{'*0} \rightarrow \Xi_b^0 \pi^0$		&0.46*&\\
 \hline

				&		&$\Xi_b^{'*-} \rightarrow \Xi_b \pi$			&1.65&\\ 
\cline{3-4}
$\Xi_b^{-}(5955)$	&$3/2^+$	&$\Xi_b^{'*-} \rightarrow \Xi_b^0 \pi^-$		&1.13*&0.79\\ 

				&		&$\Xi_b^{'*-} \rightarrow \Xi_b^- \pi^0$		&0.52*&\\
\hline\hline
  \end{tabular}
  \label{SHBG2parameter}
\end{table}

The numerical results of $\tilde{G}_1$ and $\tilde{G}_2$ are summarized in Tables \ref{SHBG1parameter} and \ref{SHBG2parameter}. As the same as the diquark model, the coupling constants of the SHB model, $\tilde{g}_1$ and $\tilde{g}_2$, are given by the spin average of $\tilde{G}_1$ and the isospin average of $\tilde{G}_2$ with the parameter $\alpha\simeq1.73$ from the assumption $g_s=3$. Their results are
\begin{eqnarray}
&&(\tilde{G}_1^c, \tilde{G}_2^c; \tilde{g}_1, \tilde{g}_2)=(0.69, 0.65; 0.61,  0.05    ),~~~
\label{g1g2CSHB}\\
&&(\tilde{G}_1^b, \tilde{G}_2^b; \tilde{g}_1, \tilde{g}_2)=(0.63, 0.74; 0.89,  -0.15    ),~~~
\label{g1g2BSHB}
\end{eqnarray} 
whose values are much different from those in the diquark model, Eqs. (\ref{g1g2CD}) and (\ref{g1g2BD}). However, the ratio $|\tilde{g}_1/\tilde{g}_2|$ is estimated to be $0.07$ from the charm sector~(\ref{g1g2CSHB}) and $0.17$ from the bottom sector~(\ref{g1g2BSHB}), which are similar to those obtained from the diquark model.
This result means that the effects of the $U(1)_A$ anomaly and the $SU(3)$ flavor symmetry breaking, included in the $g_2$ and $\tilde{g}_2$ terms, from singly bottom baryons are larger than that from singly charmed baryons.

\begin{figure}[b]
\centering
  \includegraphics[clip,width=1.02\columnwidth]{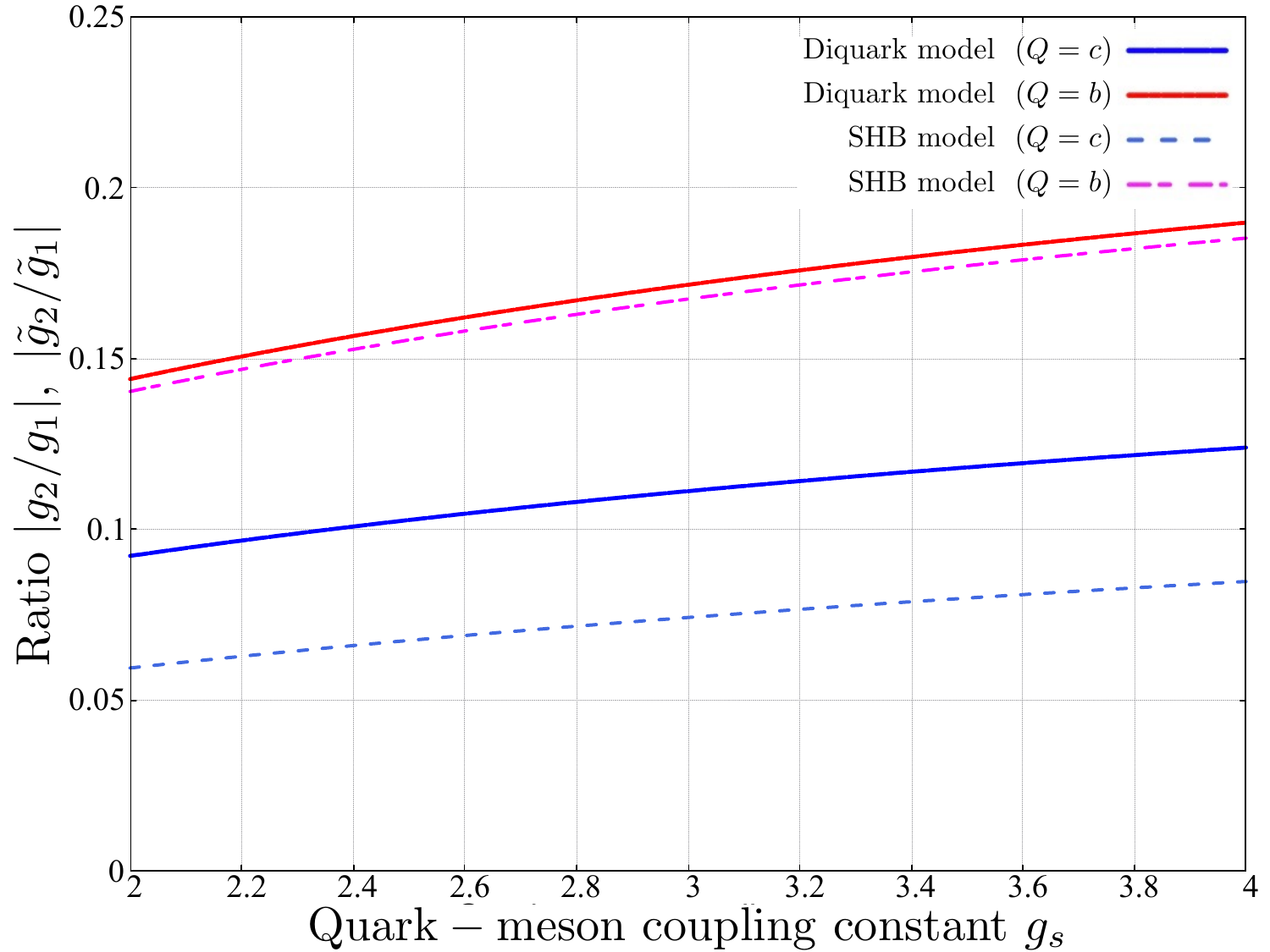}
  \caption{Dependence of the coupling constant ratio from the diquark model ($|g_2/g_1|$) and the SHB model ($|\tilde{g}_2/\tilde{g}_1|$) on the quark-meson coupling constant $g_s$. $Q=c, b$ denotes the heavy quark inside the SHBs employed as inputs.} 
  \label{figureratio}
\end{figure}
In Fig.~\ref{figureratio}, we show the ratios of coupling constants, $|g_2/g_1|$ from the diquark model in Sec.~\ref{sec:Coupling} and $|\tilde{g}_2/\tilde{g}_1|$ from the SHB model, as functions of quark-meson coupling constant $g_s$. Here, the range of $g_s$ is from 2 to 4, which corresponds to the effective $u$ or $d$ quark mass of about 200 MeV $< m_{u/d} <$ 350 MeV using the Goldberger-Treiman relation for quarks ($M_{u, d}=g_s f_\pi$).
From this figure, we find that $|g_2/g_1|$ becomes larger with increasing $g_s$ for both models.


\bibliography{diquarkdecay}
\end{document}